\documentclass[twocolumn,trackchanges]{aastex62}

\received{}
\revised{}
\accepted{}

\shorttitle{MAGIC observations of GRB 160821B}
\shortauthors{MAGIC Collaboration}

\begin{document}

\title
{MAGIC observations of the nearby short gamma-ray burst GRB~160821B}

\email{contact.magic@mpp.mpg.de}
\correspondingauthor{Koji Noda, Lara Nava, Susumu Inoue}

\collaboration{MAGIC Collaboration}
%
\author[0000-0001-8307-2007]{V.~A.~Acciari}
\affiliation{Inst. de Astrof\'isica de Canarias, E-38200 La Laguna, and Universidad de La Laguna, Dpto. Astrof\'isica, E-38206 La Laguna, Tenerife, Spain}
\author[0000-0002-5613-7693]{S.~Ansoldi}
\affiliation{Universit\`a di Udine and INFN Trieste, I-33100 Udine, Italy}
\author[0000-0002-5037-9034]{L.~A.~Antonelli}
\affiliation{National Institute for Astrophysics (INAF), I-00136 Rome, Italy}
\author[0000-0001-9076-9582]{A.~Arbet Engels}
\affiliation{ETH Z\"urich, CH-8093 Z\"urich, Switzerland}
\author{K.~Asano}
\affiliation{Japanese MAGIC Group: Institute for Cosmic Ray Research (ICRR), The University of Tokyo, Kashiwa, 277-8582 Chiba, Japan}
\author[0000-0002-2311-4460]{D.~Baack}
\affiliation{Technische Universit\"at Dortmund, D-44221 Dortmund, Germany}
\author[0000-0002-1444-5604]{A.~Babi\'c}
\affiliation{Croatian MAGIC Group: University of Zagreb, Faculty of Electrical Engineering and Computing (FER), 10000 Zagreb, Croatia}
\author[0000-0002-1757-5826]{A.~Baquero}
\affiliation{IPARCOS Institute and EMFTEL Department, Universidad Complutense de Madrid, E-28040 Madrid, Spain}
\author[0000-0001-7909-588X]{U.~Barres de Almeida}
\affiliation{Centro Brasileiro de Pesquisas F\'isicas (CBPF), 22290-180 URCA, Rio de Janeiro (RJ), Brazil}
\author[0000-0002-0965-0259]{J.~A.~Barrio}
\affiliation{IPARCOS Institute and EMFTEL Department, Universidad Complutense de Madrid, E-28040 Madrid, Spain}
% ...


\begin{abstract}
The coincident detection of GW170817 in gravitational waves and electromagnetic radiation spanning the radio to MeV gamma-ray bands provided the first direct evidence that short gamma-ray bursts (GRBs) can originate from binary neutron star (BNS) mergers.
On the other hand, the properties of short GRBs in high-energy gamma rays are still poorly constrained, with only $\sim$20 events detected in the GeV band, and none in the TeV band.
GRB~160821B is one of the nearest short GRBs known at $z=0.162$.
Recent analyses of the multiwavelength observational data of its afterglow emission revealed an optical-infrared kilonova component, characteristic of heavy-element nucleosynthesis in a BNS merger.
Aiming to better clarify the nature of short GRBs, this burst was automatically followed up with the MAGIC telescopes, starting from 24 seconds after the burst trigger.
Evidence of a gamma-ray signal is found above $\sim$0.5 TeV at a significance of $\sim3\,\sigma$ during observations that lasted until 4 hours after the burst.
Assuming that the observed excess events correspond to gamma-ray emission from GRB 160821B, 
in conjunction with data at other wavelengths, we investigate its origin in the framework of GRB afterglow models.
The simplest interpretation with one-zone models of synchrotron-self-Compton emission from the external forward shock has difficulty accounting for the putative TeV flux.
Alternative scenarios are discussed where the TeV emission can be relatively enhanced.
The role of future GeV-TeV observations of short GRBs in advancing our understanding of BNS mergers and related topics is briefly addressed.
\end{abstract}




\section{Introduction} 
\label{sec:intro}

Gamma-ray bursts (GRBs) are brief but extremely luminous flashes of radiation that occur at cosmological distances.
Their prompt emission is observed primarily as keV-MeV photons with durations ranging from milliseconds to minutes.
This is accompanied by afterglow emission that fades more gradually over timescales of hours to months and covers a much broader range of wavelengths compared to the prompt emission.
The prompt emission is believed to arise from transient, ultrarelativistic jets triggered by cataclysmic events involving neutron stars or stellar-mass black holes.
The nature of the afterglow is well understood as non-thermal emission produced by electrons accelerated in external shocks, driven by the interaction of the jet with the ambient medium (\citealt{Kumar15} and references therein).

Although GRB afterglows have frequently been observed to span the radio to GeV bands, they had eluded detection in TeV gamma rays for a long time, despite numerous searches over many decades.
A detection in the TeV band was finally achieved for GRB~190114C with the Major Atmospheric Gamma Imaging Cherenkov (MAGIC) telescopes, starting from $\sim$60 seconds after the burst in the energy range 0.2-1 TeV and beyond, which provided the first strong evidence for inverse Compton emission from the afterglow \citep{MAGIC-190114C-GCN, MAGIC-190114C-a, MAGIC-190114C-b}, as well as new constraints on Lorentz invariance violation \citep{MAGIC-190114C-c}.
The detection of gamma rays with energies above 0.1 TeV with the High Energy Stereoscopic System (H.E.S.S.) telescopes was later reported for GRB~180720B from $\sim$10 hours after the burst \citep{HESS19}, and GRB~190829A from $\sim$4 hours after the burst \citep{HESS19-2}.

The duration and spectra of GRB prompt emission exhibit a bimodal distribution that indicates two different classes of events.
With $T_{90}$ denoting the time interval containing 90\% of the prompt photon counts, 
long GRBs with durations $T_{90} \gtrsim$ 2 s, which include GRB~190114C and GRB~180720B, are widely acknowledged to be generated during the core collapse of massive stars \citep{Woosley06}.
The origin of short GRBs with $T_{90} \lesssim$ 2 s has been less certain. Mergers of binary neutron stars (BNS) were long suspected 
and supported by circumstantial evidence \citep{Berger14}.
An infrared excess observed in the afterglow of GRB 130603B, a short GRB at $z=0.356$, was interpreted as emission from a kilonova (or macronova, hereafter simply kilonova) \citep{Tanvir13}, a distinctive signature of a BNS merger powered by associated r-process nucleosynthesis of heavy elements \citep{Metzger19}
\footnote{Candidate kilonovae have also been found in retrospective searches in past short GRBs \citep{Jin18,Jin20,Rossi20}.}.
However, strong evidence for a BNS origin of short GRBs was lacking until recently.
Decisive progress occurred with the discovery of GW170817 in gravitational waves, in coincidence with GRB 170817A and the optical-infrared transient AT2017gfo, ascertained to be a kilonova \citep{Abbott17a}.
Together with radio evidence for a collimated outflow \citep{Mooley18, Ghirlanda19}, these observations provided the first strong indication that a BNS merger indeed triggers a short GRB (\citealt{Nakar19} and references therein).

Nevertheless, the properties of short GRBs remain much less understood compared to long GRBs, particularly their emission at energies above the GeV band.
Of the 186 GRBs detected by the Large Area Telescope aboard the {\it Fermi} Gamma-Ray Space Telescope ({\it Fermi}-LAT; \citealt{LAT-instrument}) from August 2008 until August 2018, only 17 are short GRBs \citep{Ajello19}.
Of the latter, only GRB 090510, a bright event at $z=0.903$, has a measured redshift \citep{Ackermann10}.
No detection of TeV-band gamma rays from a short GRB has been reported to date.

GRB 160821B is a short GRB discovered by the Burst Alert Telescope of the Neil Gehrels {\it Swift} Observatory ({\it Swift}-BAT; \citealt{Swift, BAT}) and the {\it Fermi} Gamma-ray Burst Monitor ({\it Fermi}-GBM; \citealt{GBM-instrument}). 
It is identified with a host galaxy at $z=0.162$, making it one of the nearest short GRBs known.
Recent analysis and modeling of the multiwavelength data of this GRB covering the radio to X-ray bands by two independent groups revealed good evidence for a kilonova superposed on its non-thermal afterglow emission \citep{Troja19, Lamb19}.
So far it is the best sampled kilonova without a gravitational wave detection.
However, while both groups agree on the presence of a kilonova, the detailed interpretation and inferred properties of the non-thermal afterglow as well as the kilonova differ quite significantly between the two groups.

Aiming to better understand the properties of short GRBs at energies above a few tens of GeV, follow-up observations of GRB 160821B were conducted with MAGIC telescopes. 
The low redshift of the burst is particularly important at these energies, as it mitigates the effect of photon attenuation due to $\gamma\gamma$ interactions with the extragalactic background light (\citealt{Dwek13} and references therein).
As the limited field of view of Cherenkov telescopes such as MAGIC preclude finding GRBs on their own, the standard strategy is automated follow-up of GRBs that are identified, localized and alerted by wide-field satellite instruments, which entails a time delay until the start of the observations.
Within the MAGIC GRB follow-up program \citep{Berti19}, the observation of GRB 160821B automatically started 24 seconds after the burst trigger, the shortest delay realized so far (13 seconds until MAGIC received the alert plus 11 seconds for the response of the telescopes).

An excess of gamma rays is found at the GRB position above $\sim$ 500 GeV during the observations that continued until 4 hours after the burst.
This paper reports the results of these observations, together with interpretations of the multiwavelength data based on detailed numerical modeling of the non-thermal afterglow emission.

Section 2 presents an overview of the observations of this GRB with MAGIC, {\it Fermi} and other facilities.
Section 3 describes the results of the data analysis for {\it Fermi}-LAT and MAGIC.
Section 4 discusses the theoretical interpretations of these results, in combination with multiwavelength data.
We summarize in Section~5.

\section{Observations}
\label{sec:observations}

\subsection{Radio to GeV Gamma-ray Observations}
\label{sec:grb160821b}
GRB~160821B triggered the {\it Swift}-BAT detector at 22:29:13 UT on 21 August 2016 (hereafter $T_0$; \citealt{Swift-160821B}).
With the reported burst duration $T_{\rm 90} = 0.48{\rm\,s}$, the event is classified as a short GRB. The spectrum of the prompt emission in the keV-MeV range is described by a power-law with an exponential high-energy cutoff, with photon index $0.11 \pm 0.88$, peak energy $ E_{\rm p} = (46.3 \pm 6.4)$\,keV and fluence $S(15-150{\rm\,keV})= (1.0 \pm 0.1)\times 10^{-7} ~{\rm erg~cm}^{-2}$ \citep{BATref-160821B}.
The prompt emission was also detected by {\it Fermi}-GBM at the same trigger time as {\it Swift}-BAT, with $T_{\rm 90} (50-300\,{\rm keV})\sim1\,{\rm s}$ (\citealt{GBM-160821B}; refined later to $1.088 \pm 0.977 \,{\rm s}$; \citealt{GBM-catalog}.)
The spectrum is fit with a cutoff power-law function
\footnote{https://heasarc.gsfc.nasa.gov/W3Browse/fermi/fermigbrst.html}
with $E_{\rm p} = 92 \pm 28$\,keV and fluence $S(10-1000\,{\rm keV}) = (1.7 \pm 0.2)\times 10^{-7} \,{\rm erg\,cm}^{-2}$.
A host galaxy was identified \citep{NOT-160821B}
with spectroscopic redshift $z=0.162$ \citep{WHT-160821B,Troja19,Lamb19}, making this one of the nearest short GRBs to date.
With this redshift, the isotropic energy is estimated to be $E_{\rm iso} \sim 1.2 \times 10^{49}$\,erg, which is toward the low end of the known distribution for short GRBs, but not unusual \citep{Berger14}. 

At the time of the GBM trigger, the burst was near the border of the standard field of view (FoV) of {\it Fermi}-LAT ($<$\,60 degrees). 
No emission was detected by LAT in the energy range 0.3-3 GeV (See \ref{sec:lat_results} for more details).

\begin{figure}[ht]
  \resizebox{\hsize}{!}{\includegraphics{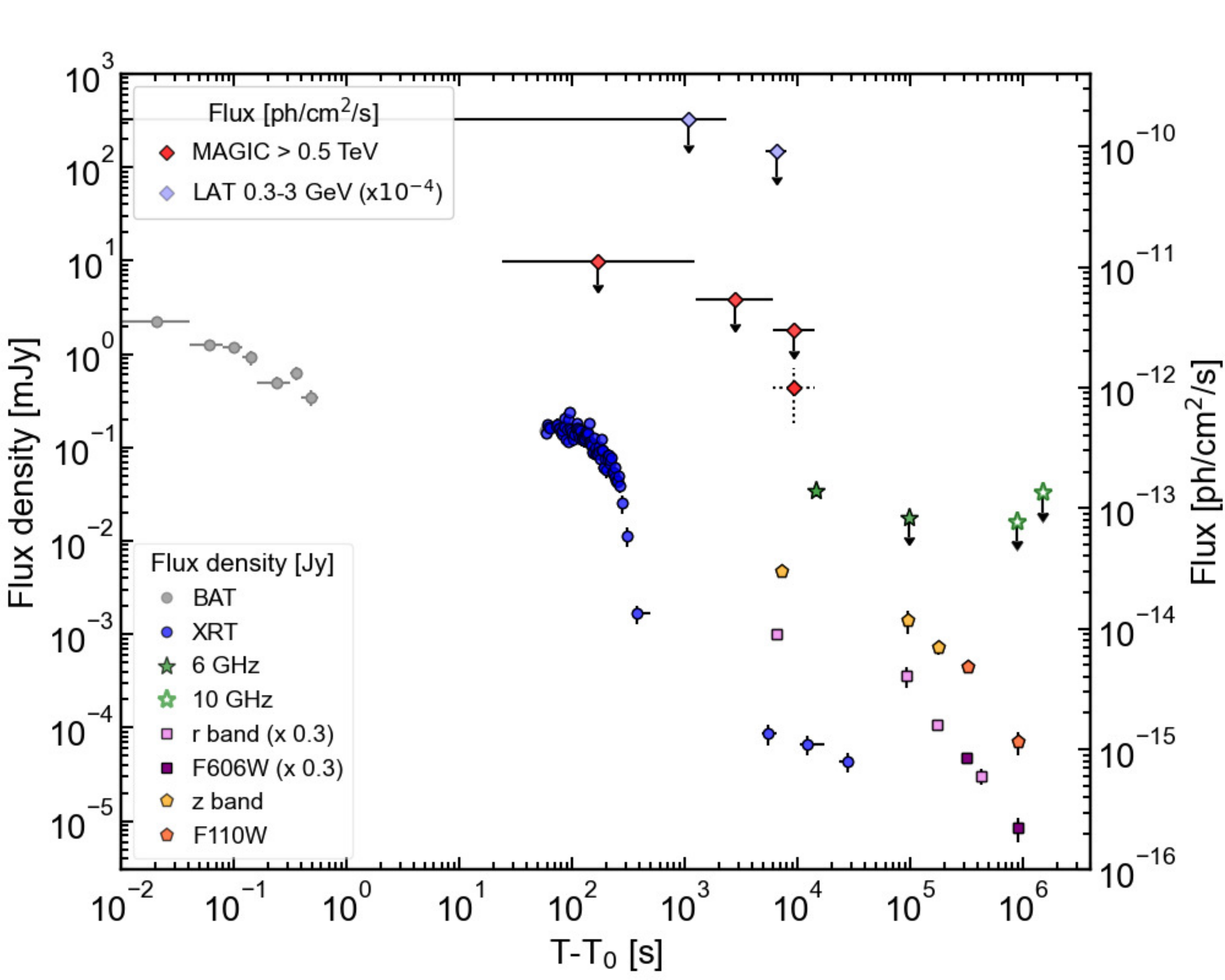}}
  \caption{Observations of GRB\,160821B at different wavelengths. Photon fluxes and flux densities are shown as a function of the observer time after the BAT trigger. The MAGIC flux is not corrected for EBL attenuation. For the third MAGIC time bin, both flux and upper limit points are shown, in view of the limited significance of the putative signal (See \ref{sec:magic_results}). LAT upper limits and $r$ band observations have been re-scaled for clarity (see legend).}
  \label{fig:lc}
\end{figure}

Follow-up observations were performed in the radio, optical and X-ray bands (see corresponding light curves in Fig.~\ref{fig:lc}).
{\it Swift} X-Ray Telescope (XRT) \citep{XRT} started observations 57\,s after the BAT trigger. The X-ray light curve, retrieved from the public on-line repository \citep{Evans09}, reveals complex behavior, with an initial plateau followed by a steep decay. 
After $\sim10^3$\,s, a more commonly observed type of decay is seen, with 
$\propto t^{-0.8}$, where $t$ is time since $T_0$ \citep{Troja19}. 

The optical afterglow was first reported by the Nordic Optical Telescope \citep{NOT-160821B}, and confirmed by the William Herschel Telescope \citep{WHT-160821B}, Gran Telescopio Canarias \citep{GTC-160821B}, and the Hubble Space Telescope \citep{HST-160821B}.
Observations at different epochs confirmed that the optical source was fading, with a reported magnitude $r = 22.6 \pm 0.1$\,mag at 0.95\,hours after $T_0$. 
The GRB is located in the outskirts of the host spiral galaxy, at $\sim$15\,kpc projected distance from its center \citep{Troja19,Lamb19}.
In the radio band, VLA (6\,GHz) detected a fading source consistent with an afterglow \citep{VLA-160821B}.

\subsection{MAGIC Observations and Data Analysis}
\label{sec:magic_data}

MAGIC is a system composed of two imaging air Cherenkov telescopes, both with a mirror diameter of 17~m.
The system is located at 2200 m above sea level, at the Roque de Los Muchachos Observatory in La Palma, Canary Islands, Spain.
The integral sensitivity of the system is 0.66\% of the Crab Nebula flux above 220 GeV with a 50-hour observation \citep{PerformancePaperII}.

MAGIC started observing GRB~160821B at the {\it Swift}-BAT position (RA: +18h 39m 57s;  Dec: +62d 23m 34s) on 21 August 2016, 22:29:37\,UT, 24 seconds after 
$T_0$. 
The observation started from a zenith angle of 34 degrees, and continued until 4 hours after $T_0$ (22 Aug, 2:29 UTC), reaching a zenith angle of 55 degrees.
The level of the night sky background (NSB) light was relatively high, due to the presence of the Moon.
The NSB quickly increased during the observations as the Moon rose, 
ranging from 2 to 8 times the level during dark nights \citep{MoonPerformance}.
The first $\sim$1.7 hours of the data (until $\sim$0:10 UTC) were strongly affected by clouds, 
while the remaining $\sim$2.2 hours were taken under better weather conditions. 
The atmospheric transmission was measured using a LIDAR facility installed at the MAGIC site \citep{Atmohead14}.
In the first part of the observation, the transmission at a height of 9 km
fluctuated between 40\%-70\% (average $\sim$60\%) relative to that in good weather conditions, which prevented the use of standard analysis procedures. 
On the other hand, this was above 85\% during the latter part, where the data quality was good enough for standard analysis with corrections for the transmission. 
Since the data for the first hours are potentially crucial for clarifying the physics of GRBs, we applied a dedicated analysis to recover this data, as described below.

The data analysis was performed entirely with the MAGIC standard analysis package called MARS \citep{MARS-2013}. 
The telescope performance, such as the sensitivity, energy threshold, and systematic errors for observations performed under nominal conditions can be found in \citet{PerformancePaperII}.
In this analysis, however, we used dedicated software configurations optimized for high NSB levels.
The description and performance study of this method can be found in \citet{MoonPerformance}.
After calibration of the data, we applied a more stringent image cleaning procedure compared to standard ones to remove a larger amount of spurious signals. As a consequence, the energy threshold is increased. 
We used data from known, bright gamma-ray sources (Crab Nebula, Mrk~421) observed under similar conditions (NSB and atmospheric transmission) to optimize the analysis cuts, in accordance with expected changes in the shower image parameters.  
The best sensitivity was obtained with a cut corresponding to an energy of $\sim$0.8 TeV. Thus, we used this threshold in order to maximize the sensitivity and search for possible signals.

\section{Results}
\label{sec:results}

\subsection{{\it Fermi}-LAT Results}
\label{sec:lat_results}

The data were selected from a region of interest (RoI) of 12 degrees centered on the best GRB position provided by {\it Swift}-XRT. 
We analyzed the data up to 10000 seconds after $T_0$, 
considering a zenith angle cut of 100 degrees to avoid Earth limb photons.
A non-standard FoV limit of $\sim$70 degrees (in contrast to the normal 60 degrees) was chosen in order to account for all analyzable data, including the entire RoI of 12 degrees at the earliest times, when GRB~160821B was at a boresight angle of $\sim$61 degrees.
The source went outside the {\it Fermi}-LAT FoV at $T_{0}$ + 2315 s. During this first interval,
the source was mainly around the border of the FoV of the instrument due to a previous automatic re-pointing request in the direction of GRB~160821A~\citep{2016GCN.19403....1L, 2016GCN.19413....1L}. 
It later reentered the FoV from $T_0$ + 5285 s up to $T_0 + 8050$ s, following the standard survey mode, moving in the LAT FoV from 70 degrees down to 10 degrees with respect to the boresight.

No hints of a detection were registered, 
and upper limits were derived adopting a Bayesian approach.
Pass8 Source data \citep{pass8} in the energy range 0.3 - 3\,GeV were selected and analysed with the unbinned likelihood method, similar to that for the Second LAT GRB Catalog \citep{secondLATGRB}, using the FermiTools package version 1.2.1\footnote{Data and software are available from the {\it Fermi} Science Support Center \url{https://fermi.gsfc.nasa.gov/ssc/data/}}, 
and the corresponding instrument response functions {\it P8R3\_SOURCE\_V2}.
The GRB was modeled as a point-like source, having a fixed power-law spectrum with photon index -2. The spectral parameters of other sources were kept fixed to those derived from the 4FGL catalog \citep{4FGL}. 
For the background modeling, appropriate isotropic extragalactic and Galactic models {\it iso\_P8R3\_SOURCE\_V2\_v1.txt}, {\it gll\_iem\_v07.fits}\footnote{The background models are available at \url{https://fermi.gsfc.nasa.gov/ssc/data/access/lat/BackgroundModels.html}} were employed.
Two main time intervals were considered during which the source was in the FoV, from $T_0$ to $T_0 + 2315$ s and from $T_0$ + 5285 s to $T_0 + 8050$ s.
The resulting upper limits in the 0.3 - 3\,GeV energy band are respectively $1.7\times 10^{-6} ~{\rm cm^{-2} s^{-1}}$ 
and $9.0\times 10^{-7} ~{\rm cm^{-2} s^{-1}}$.

\subsection{MAGIC Results}
\label{sec:magic_results}

\begin{figure}[ht]
  \resizebox{\hsize}{!}{\includegraphics{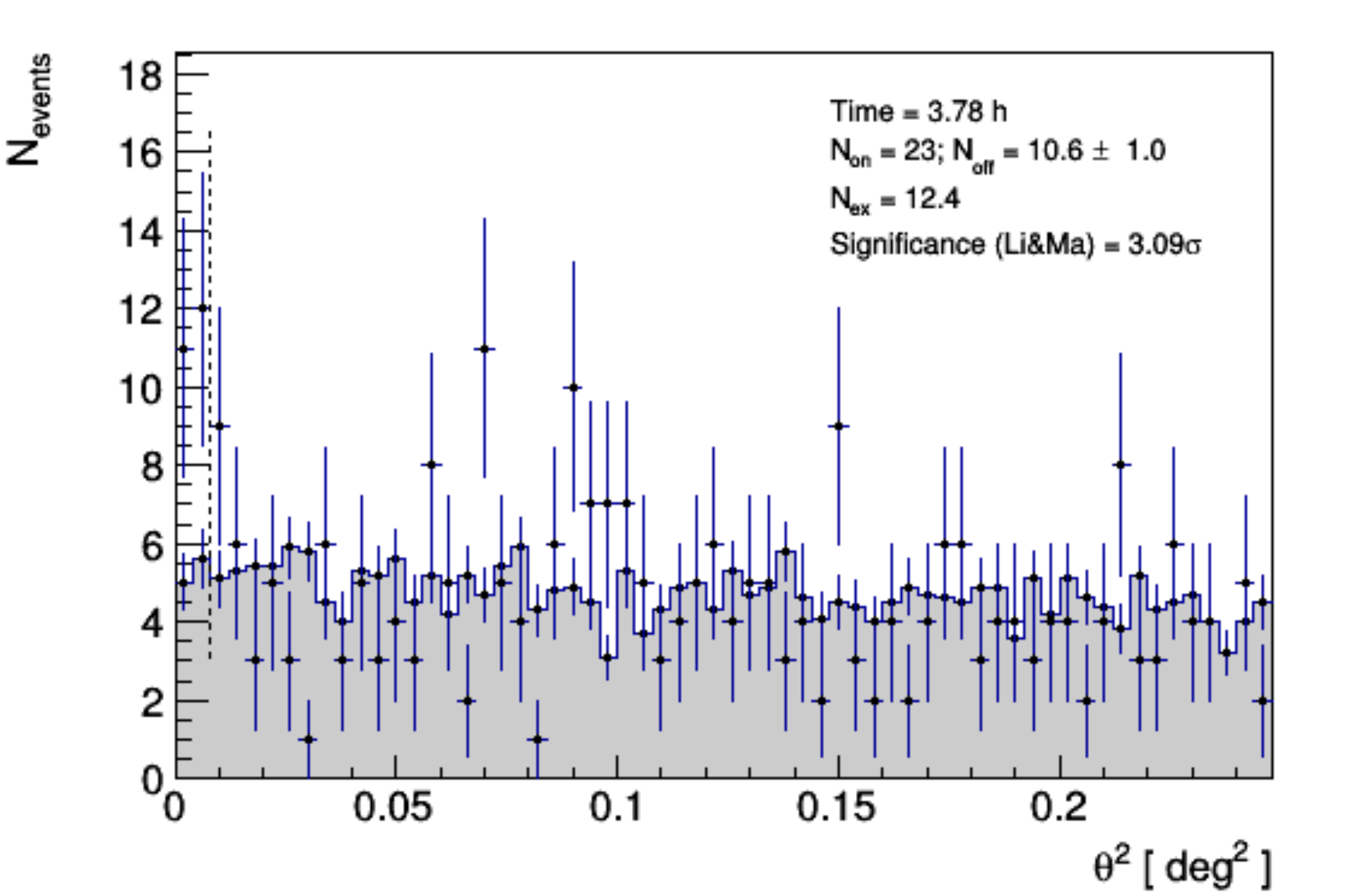}}
  \caption{Distribution of the squared angular distance $\theta^2$ between the reconstructed arrival directions of individual events and the nominal source position of GRB 160821B from {\it Swift}-XRT (data points),
  and that estimated at the background positions (10 regions; shaded area).
  Statistical uncertainties on the number of events are shown as vertical error bars.
  The number of excess events (Nex = 12.4 +/- 4.9) and the significance are evaluated for the range between 0 and the vertical dashed line. The estimated energy threshold is $\sim$800 GeV. All data taken for GRB~160821B on the night of 21 August 2016 were used.}
  \label{fig:th2}
\end{figure}

The excess significance at the GRB position observed by MAGIC is 3.1 $\sigma$ (pre-trial). 
It is derived using the prescription of eq. 17 of \citet{LiMa}, based on the distribution of the squared angular distance ($\theta^2$) between the reconstructed source position and the nominal source position (Fig. \ref{fig:th2}). 
We tested two analysis cuts used for MAGIC data, the cut described above and an alternative cut that is optimized for low energy events, so that a trial factor of 2 should be considered, leading to a corrected significance value of 2.9 $\sigma$. 

\begin{figure}[ht]
  \resizebox{\hsize}{!}{\includegraphics{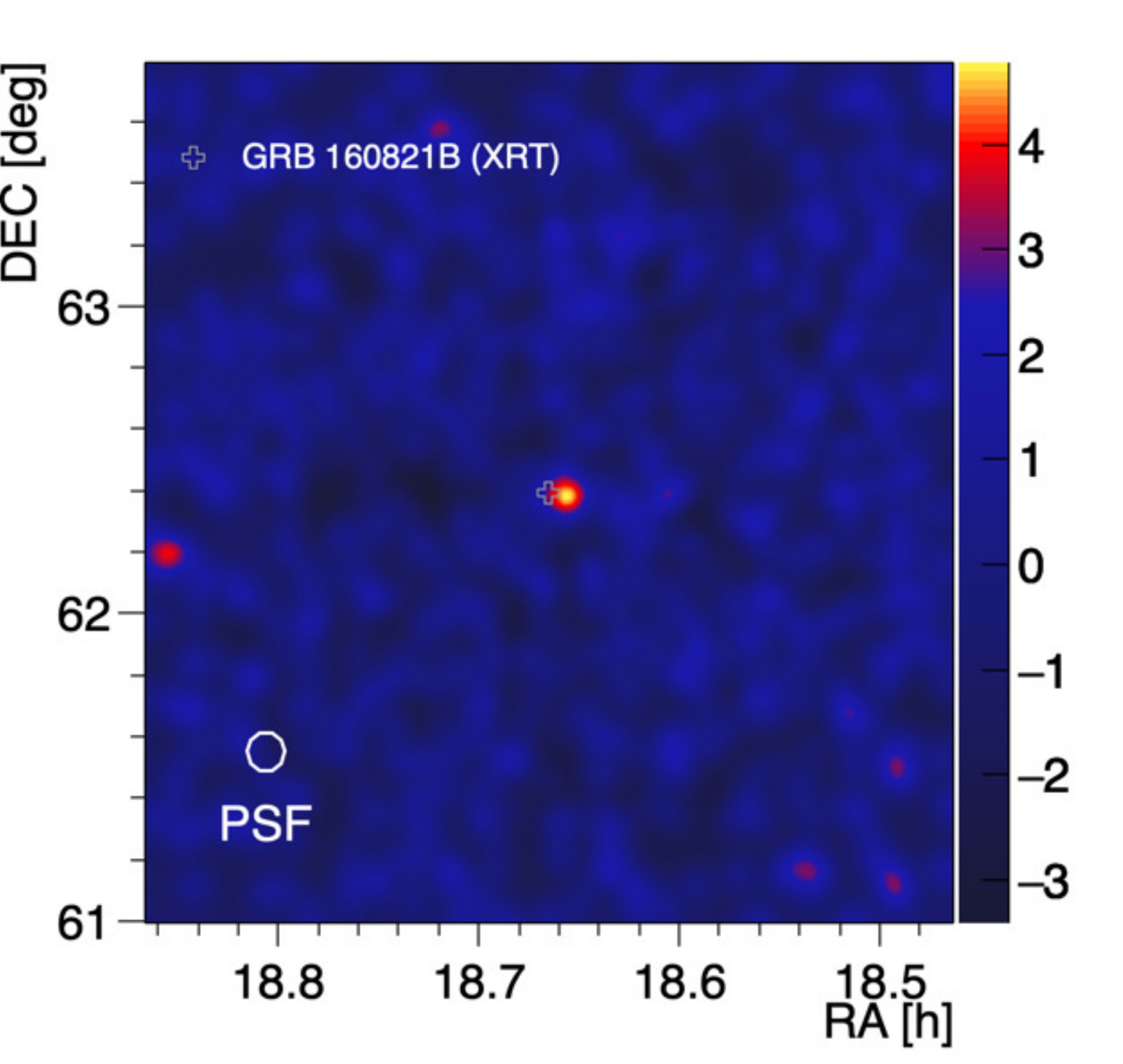}}
  \caption{Sky map showing the excess significance (standard deviation, pre-trial) as measured by MAGIC for events above $\sim$ 0.8 TeV. The white cross marks the position of GRB~160821B according to {\it Swift}-XRT. 
  The PSF corresponding to 68\% containment is depicted as a white circle in the left lower corner, with radius 0.045 deg.}
  \label{fig:skymap}
\end{figure}

We also computed a significance sky map of the observation (Fig. \ref{fig:skymap}) and found a spot with high significance (4.7 $\sigma$ pre-trial) 0.05 deg away from the GRB position. 
The post-trial significance of seeing such a hot spot at any place in the sky map is 3.0 $\sigma$ (1232 trials). 
Since this hot spot is close to the GRB position, we discuss whether it can be a possible signal from the GRB that appears displaced from its actual position. 

The systematic error in the telescope pointing is typically $<$0.02 degrees and maximally $\sim$0.03 degrees even with strong wind gusts.
Thus the offset of 0.05 degrees cannot be attributed to the telescope pointing alone. 
We also checked in the 4FGL catalog \citep{4FGL} that there are no previously known GeV gamma-ray sources within 1 deg around the spot that could be potential TeV emitters. 

We considered possible shifts of the reconstructed source position for a weak source embedded in a background that is fluctuating at a comparable level.
We performed a Monte Carlo study simulating the sky maps, and found  
that the centroid position of the hot spot can be spread over a larger area than that of the actual signal.  
The hot spot position is distributed as a 2-dimensional Gaussian with a width 2.6 times larger than that of the signal. 
The probability of the reconstructed position of such weak sources falling outside the original 1-$\sigma$ contour of the point spread function (PSF, 0.045 deg in radius) is 24\%. 
Therefore we conclude that the 0.05-degree offset seen in the skymap is well explained as statistical fluctuations in the case of weak signals, 
and that the significance of 3.1 $\sigma$ (pre-trial) conservatively computed at the {\it Swift}-XRT position can be regarded as evidence of a signal from the GRB. 

We note that in addition to the trial factor discussed above, follow-up observations of other GRBs in the MAGIC GRB program may be considered as further trials. 
Among the 69 GRBs followed up by MAGIC in stereoscopic mode since 2009 \citep{MAGIC-GRB_ICRC15,MAGIC-GRB13_15}, the only short GRBs other than GRB 160821B observed under acceptable conditions were 140930B, 160927A, and 180715A, all with delays longer than 5000 seconds, and none with measured redshifts.
Properly accounting for such observations as trials is difficult and not discussed in this paper, as they are subject to hidden observational and analysis biases, implying unequal trial factors.

In order to estimate flux values, we divided the data into two sets according to the weather conditions during the observations. 
The first 1.7 hours are characterized by low atmospheric transmission (average $\sim$60\%), while the remaining 2.2 hours had good weather conditions. The first 1.7 hours are further subdivided into two time bins, to better represent the results on a logarithmic time scale. 
The resulting bins in time since $T_0$ are $24$ s to $1216$ s, $1258$ s to $6098$ s, and $6134$ s to $14130$ s. 
The flux is estimated by integrating the signal above 0.5 TeV, 
the peak energy of the reconstructed gamma rays when assuming a power-law spectrum with photon index -2, convolved with the effective area.
Because of the low significance in the first two time bins, we calculated 95\% confidence-level flux upper limits using the method described in \citet{Rolke}, obtaining 1.1$\times 10^{-11} ~{\rm cm^{-2} s^{-1}}$ and 5.4$\times 10^{-12} ~{\rm cm^{-2} s^{-1}}$, respectively.
For the third time bin, we can similarly derive a flux upper limit of $3.0 \times 10^{-12} ~{\rm cm^{-2} s^{-1}}$.
On the other hand, despite the limited significance, we can also derive the flux for the last time bin, assuming that the excess is a real signal, which gives
$9.9 \pm 4.8 \times 10^{-13} ~{\rm cm^{-2} s^{-1}}$. 

In order to check for the possibility of an unknown, unrelated gamma-ray source at the GRB position, we carried out an additional observation about a year after the GRB (11-14 Sep 2017, $T_0 + 3.3 \times 10^7$ s) and obtained 7.6 hours of good quality data. 
The result is a flux upper limit of $4.4 \times 10^{-13} ~{\rm cm^{-2} s^{-1}}$ ($>$ 0.5 TeV, 95\% C.L.), which is about half of the value discussed above for the putative signal.
If a steady source was present at the position, an observation of 7.6 hours (instead of 2.2 hours) should result in a flux measurement with a smaller error, $9.9 \pm 2.6 \times 10^{-13} ~{\rm cm^{-2} s^{-1}}$. 
The confidence belts of the flux inferred earlier and the flux upper limit derived later marginally overlap at 2-$\sigma$ level on both sides, so the hypothesis of a steady source is disfavored, although it does not exclude the possibility of a variable source that is unrelated to the GRB. 

\begin{figure*}
  \centering
  \resizebox{1.04\hsize}{!}
  {
  \includegraphics{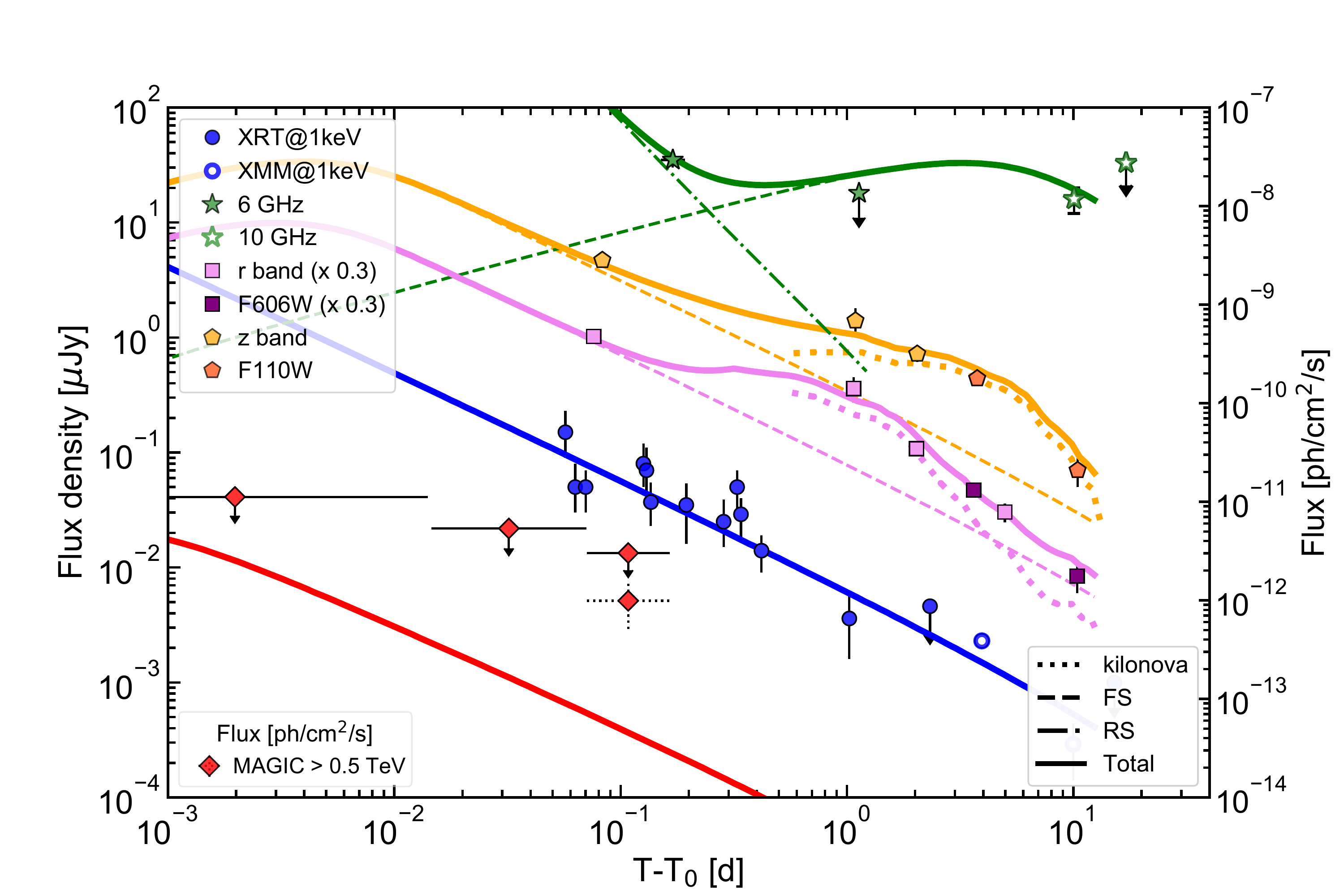}
  \includegraphics{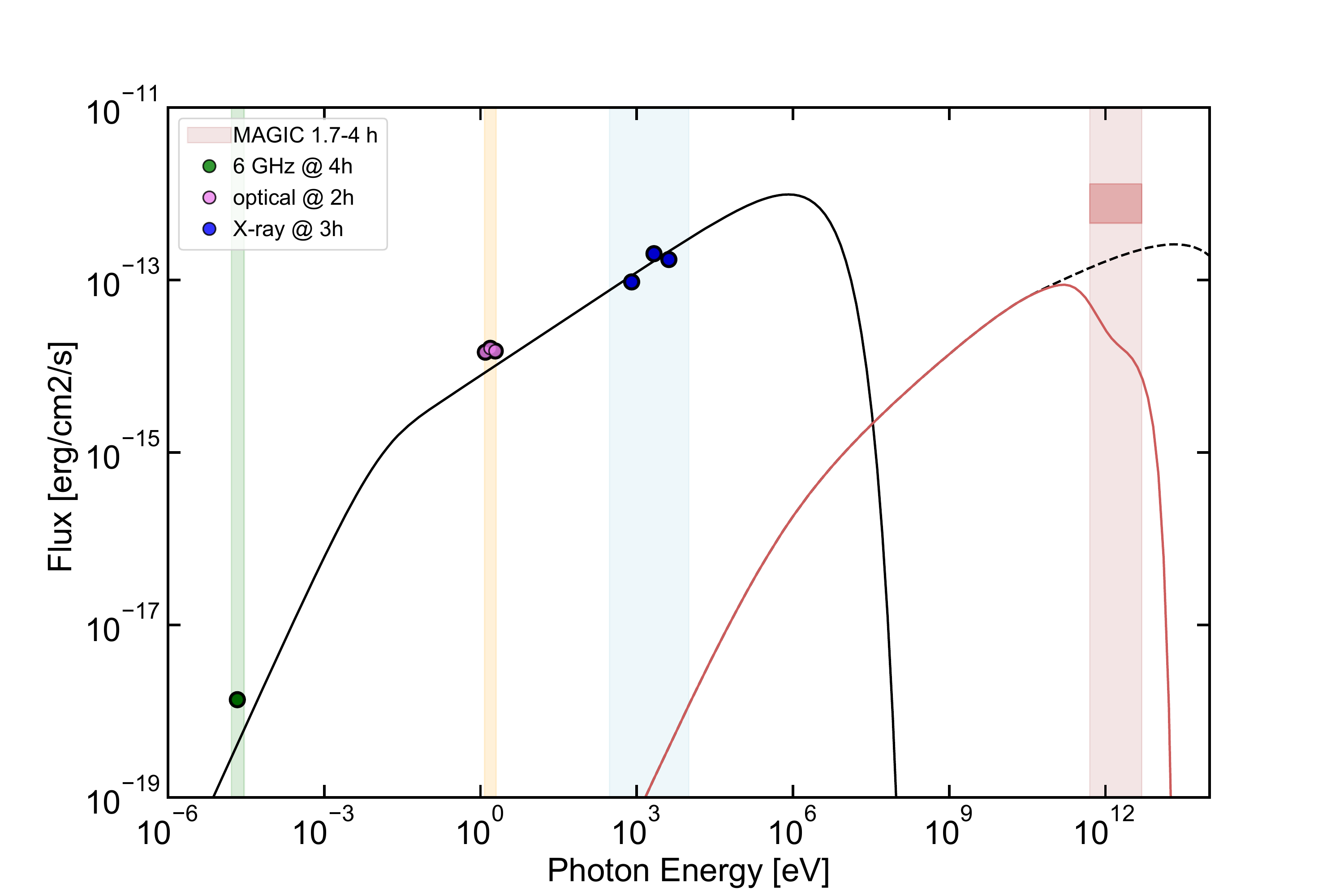}}
  \caption{Multi-wavelength data of GRB~160821B compared with afterglow modeling. The forward shock synchrotron and SSC emissions were evaluated using the following afterglow parameters: Log\,$\epsilon_{\rm e}=-0.1$, Log\,$\epsilon_{\rm B}=-5.5$, $E_{\rm k}=10^{51}$\,erg, $n = 0.05$\,cm$^{-3}$, and $p=2.2$. 
  Left: light curves at different frequencies (see legend). The modeling is shown with solid curves. The optical/nIR flux is the sum of the contribution from the forward shock (FS, dashed) and from the kilonova (dotted, from \citealt{Troja19}). The radio emission is initially dominated by the reverse shock (RS, dot-dashed, from \citealt{Troja19}). The X-rays at $t>10^3$\,s is always dominated by the forward shock.
  The red solid curve is corrected for EBL attenuation, while the MAGIC flux points are uncorrected.
  Data in the $r$ band are re-scaled for clarity (see legend).
  Right: multi-wavelength SED at approximately 3\,hours (see legend for the exact times). Shaded areas show the energy ranges covered by the instruments. 
  The thin red box only indicates the flux level measured with MAGIC and does not represent the spectral shape.
  Solid black: synchrotron emission; 
  dashed black: intrinsic SSC emission; 
  solid red: SSC emission after EBL attenuation.
  LAT upper limits are not shown, as they correspond to fluxes larger than $10^{-10}$\,erg\,cm$^{-2}$\,s$^{-1}$.}  
  \label{fig:modeling}
\end{figure*}

Because of the low significance, an unfolded spectral energy distribution could not be derived, even for the third time bin with data obtained during good weather.
The error box shown in the right panel of Fig.~\ref{fig:modeling}
indicates only the reconstructed flux for this bin, derived from the photon flux by integrating over the energy range 0.5 - 5 TeV and assuming a power-law spectrum with photon index -2 (horizontal edges of the box).
The height of the box corresponds only to statistical errors for the photon flux, and does not account for systematic errors related to the assumed spectral index.

\section{Discussion}
\label{sec:discussion}

\subsection{Modeling of Observations from Radio to X-rays}\label{syn_model}
Several distinct components can contribute to the radio to X-ray emission of short GRBs after their prompt emission.
The main component is synchrotron radiation from electrons accelerated in external forward shocks, triggered by interactions between the relativistic jet and the ambient medium (hereafter, simply ``afterglow'' radiation).
In some cases, another component can arise from a reverse shock propagating into the jet ejecta.
Two additional components are unique to short GRBs.
Often seen in X-rays is ``extended emission'', where a relatively shallow temporal decay during the first tens to hundreds of seconds is followed by a much steeper decay, widely thought to be related to long-lasting activity of the central engine (either a magnetar or a black hole resulting from a NS merger;  \citealt{Norris06, Lue15}).
Finally, optical-infrared kilonova emission can occur on timescales of days, powered by freshly synthesized r-process elements ejected in NS mergers \citep{Metzger19}.

All four of the aforementioned components are actually observed in GRB 160821B.
Hereafter, our modeling focuses on the afterglow component from the external forward shock.
Thus, we only consider the X-ray data at $t>10^3$\,s, excluding the extended emission that can be clearly seen at earlier times in Fig.~\ref{fig:lc} (see also \citealt{Zhang18}).
The kilonova emission has been inferred to dominate the optical/nIR band from 1 day to 4 days after $T_0$ \citep{Kasliwal17, Jin18, Troja19, Lamb19}.

The broad-band light curves are shown in Fig.~\ref{fig:modeling} (left-hand panel).
We adopt the X-ray light curve from \cite{Troja19} and model the broad-band emission as synchrotron emission from the external forward shock, considering the simplest case of impulsive energy injection.
The modeling is performed with a numerical code that self-consistently solves the evolution of the electron distribution, accounting for continuous electron injection with a power-law energy distribution ($dN/d\gamma\propto\gamma^{-p}$), synchrotron, synchrotron-self-Compton (SSC) and adiabatic losses, synchrotron self-absorption and $\gamma\gamma$ pair production 
(for a description of the code, see \citealt{MAGIC-190114C-b} and references therein). 

The broad-band SED at $t\sim T_0+$3\,h is shown in Fig.~\ref{fig:modeling} (right-hand panel).
The consistency between the X-ray and optical spectral indices ($F_\nu\propto\nu^{-0.8}$) suggests that the X-ray and optical bands are located between the characteristic synchrotron frequency $\nu_{\rm m}$ and the cooling frequency $\nu_{\rm c}$. 
The radio data at 6 and 10\,GHz together with optical and X-ray data constrain $\nu_{\rm m}$ to be located between the radio and optical bands. The radio emission from the forward shock is then expected to increase with time (see dashed green curve in the left-hand panel of Fig.~\ref{fig:modeling}), implying that the observed radio emission at early times is dominated by another component, most likely from the reverse shock \citep{Troja19,Lamb19,Lamb20}. To be consistent with the radio upper limits at later times, $\nu_{\rm m}$ must cross the radio band. All together, these observations constrain its value to be $\nu_{\rm m}\gtrsim 4\times10^{12}$\,Hz at $t\sim10^4$\,s and $F^{\rm syn}_{\nu_{\rm m}}\sim 0.03$\,mJy.
The model parameter space is further constrained by the requirement $\nu_{\rm c}>\nu_{\rm X}$ up to at least 4 days (from the observed lack of a clear temporal break in X-rays).
Order of magnitude estimates for the model parameters can be inferred by solving the equations
$$\nu_{\rm m}(t\sim10^4 {\rm s})\sim 2\times10^{12}\,{\rm Hz}\,\epsilon_{\rm e,-1}^2\,(p-2)^2/(p-1)^2\epsilon_{\rm B,-4}^{1/2}\,E_{\rm k,50}^{1/2} $$
$$= 4\times10^{12}\,{\rm Hz},$$
$$F^{\rm syn}_{\nu_{\rm m}}(t\sim10^4 {\rm s})\sim 0.04\,{\rm mJy}\,\epsilon_{\rm B,-4}^{1/2}\,n_{-1}^{1/2}\,E_{\rm k,50}=0.03\,{\rm mJy}, {\rm and} $$
$$\nu_{\rm c}^{\rm syn}(t\sim1\,{\rm d})\sim 4\times10^{20}\,{\rm Hz}\,\epsilon_{\rm B,-4}^{-3/2}\,n_{-1}^{-1}\,E_{\rm k,50}^{-1/2}>2.4\times10^{18}\,{\rm Hz}$$ 
(see e.g., \citealt{sari98,granotsari,pk}), where $E_{\rm k}$ is the initial, isotropic-equivalent kinetic energy, $n$ the density of the surrounding medium, $\epsilon_{\rm e}$ and $\epsilon_{\rm B}$ the fraction of energy dissipated behind the shock in accelerated electrons and the magnetic field, respectively, and $p$ the power-law index of the injected electron energy distribution. 

We find good agreement for values of the model parameters within the following ranges:
Log$(E_{\rm k}/\rm{erg})=[50-51]$, 
Log$(\epsilon_{\rm e})=[-1;-0.1]$,
Log$(\epsilon_{\rm B})=[-5.5; -0.8]$,
Log$(n/{\rm cm}^{-3})=[-4.85; -0.24]$, and
$p=[2.2;2.35]$.
The inferred values are very similar to the values inferred by \cite{Troja19}.

There is degeneracy between the parameters, that can be understood as follows: 
since $\nu_{\rm m}\propto\epsilon_{\rm e}^2\sqrt{\epsilon_{\rm B}\,E_{\rm k}}$ and 
$F_{\nu_{\rm m}}\propto E_{\rm k} \sqrt{\epsilon_{\rm B}\,n}$
for a fixed value of $\epsilon_{\rm e}$, the other parameters must satisfy $\epsilon_{\rm B}\propto E_{\rm k}^{-1}$ and $n\propto E_{\rm k}^{-1}$. 
$E_{\rm k}<10^{50}$\,erg would imply large values of $\epsilon_{\rm B}$ and $n$, resulting in $\nu_{\rm c}<\nu_{\rm X}$.

The result of the modeling is compared with observations in Fig.~\ref{fig:modeling}. The reverse shock and kilonova components (dot-dashed and dotted curves in the left-hand panel) are taken from \cite{Troja19}.

Note that in contrast to \cite{Troja19} and our modeling here, \cite{Lamb19} proposed a different, multi-zone interpretation for the afterglow, invoking emission from a narrow jet component, as well as a slower outflow component caused by energy injection from the central engine at late times. The different interpretation is mainly driven by a different analysis of the X-ray data, resulting in an X-ray light curve with evidence for a double peak.

\subsection{Modeling of the TeV Radiation}
Assuming that the TeV $\gamma$-ray signal obtained from MAGIC observations of GRB~160821B is real,
we discuss possible mechanisms for TeV emission in short GRBs and assess their viability in accounting for these observations.

\subsubsection{Synchrotron-Self-Compton emission (SSC)}
Considering the parameter space constrained from the observed synchrotron emission in the radio, optical and X-ray bands, we estimate the associated SSC component, and compare the results with MAGIC observations.
Given the wide range of values still allowed,
the expected flux of the SSC emission can vary by a few orders of magnitude.

The energy range covered by MAGIC observations lies in the range $\nu_{\rm m}^{\rm SSC}<\nu_{\rm MAGIC}<\nu_{\rm c}^{\rm SSC}$ where the flux can be analytically estimated by 
\begin{equation}
    F^{\rm SSC}(\nu)=\tau\,F^{\rm syn}(\nu_{\rm m}^{\rm syn})\,\gamma_{\rm m}^2\left(\frac{\nu}{\nu_{\rm m}^{\rm SSC}} \right)^{-(p-3)/2},
\end{equation}
subject to corrections of order $\sim\ln(\nu/\nu_{\rm m}^{\rm SSC})$, where $\tau=\sigma_{\rm T}\,R\,n/3$ and $\nu_{\rm m}^{\rm SSC}\simeq \gamma_{\rm m}^2\,\nu_{\rm m}^{\rm syn}$ \citep{SariEsin}.

The SED in the right-hand panel of Fig.~\ref{fig:modeling} shows the inferred flux in the energy range 0.5-5\,TeV under the assumption that the gamma-ray signal from GRB\,160821B is real (thin red box).
This TeV flux would imply a large amount of energy in the SSC component, with a Compton parameter $Y>1$. 
A large TeV flux is obtained for large values of $\epsilon_{\rm e}$, since they allow for lower $\epsilon_{\rm B}$ and higher $n$ (see equations in Sec.~\ref{syn_model}).
Adopting the following parameter values:
Log$(E_{\rm k}/\rm{erg})=51$, 
Log$(\epsilon_{\rm e})=-0.1$,
Log$(\epsilon_{\rm B})=-5.5$,
Log$(n/{\rm cm}^{-3})=-1.3$, and
$p=2.2$,
at the time of the last MAGIC observation, the external shock radius is $R\sim2.4\times10^{17}$\,cm, the bulk Lorentz factor is $\Gamma\sim16$, and the characteristic Lorentz factor of the electrons is $\gamma_{\rm m}\sim 3.8\times10^3$.
The expected SSC flux at 1\,TeV is then $F^{\rm SSC}(1{\rm \,TeV})\sim2\times10^{-13}$\,erg\,cm$^{-2}$\,s$^{-1}$.
The expected SSC spectrum accounting for EBL attenuation \citep{Dominguez11} is shown as the red solid curve in Fig.~\ref{fig:modeling}.
Compared with the flux suggested from MAGIC observations at 0.1 days, the maximum flux expected in one-zone SSC models falls short by about an order of magnitude.

\subsubsection{Proton synchrotron emission}
Synchrotron emission by protons accelerated to ultrahigh energies in the external shock has also been proposed as a mechanism for GeV-TeV emission in GRB afterglows \citep{Vietri97,Zhang01},
including the bright GeV emission observed in GRB 090510 \citep{Razzaque10}.
We discuss its viability for the putative TeV emission of GRB 160821B, following the analytic formulation of \citet{Zhang01}.

The maximum expected energy of proton synchrotron emission in the observer frame is
$\varepsilon_{\rm psyn,max} = 0.031\, {\rm GeV}\, \eta^{-2}\, \epsilon_B^{3/2}\, (n_0\, E_{\rm k,51})^{3/4}\, t_h^{-1/4}\, (1+z)^{-3/4} $,
where $E_{\rm k}=10^{51}\, E_{\rm k,51}\, {\rm erg}$,
$n=n_0 \ {\rm cm^{-3}}$,
$t_h$ is the observer time after the burst in hours,
and $\eta$ is a factor of order unity that characterizes the acceleration timescale.
Even when assuming optimistic values of $\epsilon_B=0.5$ and $\eta=1$, realizing $\varepsilon_{\rm psyn,max} \gtrsim 0.5$ TeV at $t \sim 2$ h for a GRB at $z=0.162$ requires $n_0\,E_{\rm k,51} \gtrsim 6000$, much larger than typical for short GRBs ($E_{\rm k} \sim 10^{49}-10^{52}$ erg, $n_0 \sim 10^{-3}-1 \ {\rm cm^{-3}}$; \citealt{Berger14}). 
It is also inconsistent with inferences from the radio to X-ray emission discussed in \ref{syn_model}. 

The requirement to reproduce the inferred TeV-band flux is likewise severe. 
For a power-law energy distribution with index $-p_p$ for the accelerated protons, their synchrotron emission is expected to have a single power-law spectrum with photon index $\alpha_{\rm int} =-(p_p+1)/2$, from a minimum energy
$\varepsilon_m = 1.7 \times 10^{-9}\, {\rm eV}\, \xi_p^{-2}\, \epsilon_p^2\, \epsilon_B^{1/2}\, E_{\rm k,51}^{1/2}\, t_h^{-3/2}\, (1+z)^{1/2}$
up to $\varepsilon = \varepsilon_{\rm psyn,max}$,
where $\xi_p$ is the fraction in number of protons swept up by the shock that are accelerated, and $\epsilon_p$ is the fraction of energy in accelerated protons relative to that dissipated behind the shock.
Assuming $p_p = 2$ typically expected for shock acceleration,
and that the spectrum extends to at least $\varepsilon=0.5$ TeV,
the energy flux at 0.5 TeV is
$F(\varepsilon=0.5\, {\rm TeV}) = 8.8 \times 10^{-15}\, {\rm erg\, cm^{-2}\, s^{-1}}\,
\epsilon_p\, \epsilon_B^{3/4}\, n_0^{1/2}\, E_{\rm k,51}^{5/4}\, D_{28}^{-2}\, t_h^{-3/4}\, (1+z)^{5/4}$,
where $D= 10^{28}\, D_{28}\, {\rm cm}$ is the luminosity distance of the GRB.
With optimistic assumptions of $\epsilon_B=0.5$, $\eta=1$ and $\epsilon_p=0.5$,
accounting for the inferred 0.5-1 TeV flux at $t \sim 2$ h
necessitates $n_0^{1/2}\, E_{\rm k,51}^{5/4} \gtrsim 4000$, far larger than expected and contradicting the constraints from lower energy bands.
Thus, proton synchrotron emission is strongly disfavored as the origin of the TeV emission suspected in GRB 160821B.

Photohadronic cascade emission, triggered by interactions of ultrahigh-energy protons with ambient low-energy photons, is another potential GeV-TeV emission mechanism in GRB afterglows \citep{Boettcher98,Zhang01}.
However, as with proton synchrotron emission, its radiative efficiency is generally low, and is disfavored for similar reasons.

\subsubsection{Other possibilities}
The TeV emission inferred for GRB 160821B from MAGIC observations may be difficult to reproduce with hadronic emission models or the simplest, one-zone SSC emission models.
Although detailed studies are beyond the scope of this paper, below we discuss some other processes that may potentially account for the putative TeV emission. 

External Compton emission, whereby accelerated electrons Compton upscatter soft photons originating from outside the emission region, is a process that is widely discussed in the context of gamma-ray emission from blazars \citep{Madejski16}, but has received relatively little attention for GRBs (see however \citealt{Murase11}).
Potential sources of external soft photons for short GRB afterglows include the extended X-ray emission \citep{Murase18},
emission from a cocoon surrounding the jet \citep{Kimura19}, as well as the kilonova \citep{Linial19}.

It is noteworthy that in contrast to \citet{Troja19}, \citet{Lamb19} advocate a two-component jet for GRB 160821B, consisting of a fast, narrow component and an additional, slower component, to better account for observed features in the X-ray light curve.
If the geometry is such that the two components are co-axial with the slower component surrounding the faster component, synchrotron photons from the former can act effectively as external soft photons for the latter and vice-versa, analogous to certain models developed for blazars in which inverse Compton emission can be enhanced compared to one-zone models \citep{Ghisellini05}.

Besides radiation from a non-thermal electron population, \cite{Vurm17} propose that inverse Compton upscattering of X-ray afterglow photons by thermal plasma at the forward shock can produce luminous TeV emission up to several hours after the burst.
They suggest that this process can be relatively efficient for short GRBs occurring in low-density environments.

Finally, GRB 160821B shows clear evidence for a reverse shock component in the radio band \citep{Troja19,Lamb19,Lamb20}.
Emission up to TeV energies may be possible from the reverse shock due to the SSC process \citep{Wang01}, adding to the TeV emission from the forward shock.

\section{Summary and outlook} 
\label{sec:conclusions}
GRB\,160821B is a short GRB found by {\it Swift} and {\it Fermi} that occurred at $z=0.162$.
Its afterglow emission was followed up by several telescopes, resulting in detections in the radio, optical-infrared, and soft X-ray bands.
Optical-infrared observations clearly revealed the presence of a kilonova, dominating the emission at $\sim$1 day.

The MAGIC telescopes also followed up this event, starting from 24 seconds and lasting until $\sim$4 hours after the burst trigger.
Non-optimal weather and observing conditions resulted in a high energy threshold for data analysis ($>$ 0.5 TeV) and limited the effectiveness of the observations, especially at early times when the afterglow is expected to be brighter and the chances for detection of a possible TeV counterpart are higher.
Nevertheless, we obtained a TeV gamma-ray excess with a significance of 3.1 $\sigma$ (pre-trial).
We also analyzed {\it Fermi}-LAT data and derived upper limits in the GeV energy range.
Collecting radio, optical and X-ray data that are publicly available,
the MAGIC observations were compared with those at lower energies.
The estimated energy flux for the possible TeV detection after correcting for EBL attenuation is about 5-10 times larger than that measured in X-rays at the same time $t\sim3$\,h.
The data from radio to X-rays can be well described as synchrotron radiation from the forward shock, with a contribution to the radio flux at early times from the reverse shock and to the optical-UV flux from the kilonova.

Assuming that the 3.1-$\sigma$ excess is indeed a gamma-ray signal associated with GRB 160821B, we discussed some models that may potentially account for such TeV emission from afterglows of short GRBs. 
Utilizing a one-zone numerical model of synchrotron emission from the external forward shock, we computed the accompanying SSC emission. Within the parameter space constrained by radio to X-ray observations, we find that the SSC flux is maximized for large values of $\epsilon_{\rm e}\sim0.8$, very low values of $\epsilon_{\rm B}\sim 3\times 10^{-6}$, and density $n\sim0.05$\,cm$^{-3}$.
The TeV flux derived under these conditions falls short of the observationally inferred flux by about a factor of 10.
Moreover, hadronic processes such as proton synchrotron emission are strongly disfavored as the origin of the possible TeV emission due to their low radiative efficiency.
Thus, the simplest, one-zone SSC emission models or hadronic emission models are unable to account for the TeV emission of GRB 160821B indicated by the MAGIC observations.

Other possibilities for interpreting the putative TeV emission from GRB 160821B should be explored in the future, such as external Compton processes, particularly in the context of two-component jet models as favored by \cite{Lamb19}, inverse Compton upscattering by thermal plasma in the forward shock, and SSC emission from the reverse shock. 

Finally, it is worth emphasizing that observations in the radio to X-ray bands alone are not necessarily sufficient for clearly characterizing short GRB afterglows, which comprise not only the nonthermal emission from the external forward shock, but also additional components including the reverse shock, extended emission, and the kilonova, all of which are seen in GRB 160821B.
Through a detailed study of one-zone models, we have demonstrated that reproducing the available data from radio to X-rays still leaves a considerable degeneracy in the range of allowed afterglow parameters. 
It also remains uncertain whether multi-zone considerations \citep{Lamb19} are needed.
In turn, this can leave fairly significant ambiguities in the inferred properties of the superposed kilonova, such as the mass, velocity and composition of the ejecta.
Our modeling also shows that while staying consistent with the observed radio to X-ray data, the SSC flux at higher energies can vary by a few orders of magnitude, depending on the afterglow parameters. 
Future GeV-TeV observations of inverse Compton emission and other emission components beyond the synchrotron component in short GRB afterglows should play an important role in disentangling such uncertainties, and advancing our knowledge of the nature of short GRBs, BNS mergers, and the origin of heavy elements.

\acknowledgments
We are grateful to Nial Tanvir and Gavin Lamb for valuable discussions.
Pierre Colin is acknowledged for his crucial contributions to the initial stages of this work.

We would like to thank the Instituto de Astrof\'{\i}sica de Canarias for the excellent working conditions at the Observatorio del Roque de los Muchachos in La Palma. The financial support of the German BMBF and MPG; the Italian INFN and INAF; the Swiss National Fund SNF; the ERDF under the Spanish MINECO (FPA2017-87859-P, FPA2017-85668-P, FPA2017-82729-C6-2-R, FPA2017-82729-C6-6-R, FPA2017-82729-C6-5-R, AYA2015-71042-P, AYA2016-76012-C3-1-P, ESP2017-87055-C2-2-P, FPA2017-90566-REDC); the Indian Department of Atomic Energy; the Japanese ICRR, the University of Tokyo, JSPS, and MEXT; the Bulgarian Ministry of Education and Science, National RI Roadmap Project DO1-268/16.12.2019 and the Academy of Finland grant nr. 320045 is gratefully acknowledged. This work was also supported by the Spanish Centro de Excelencia ``Severo Ochoa'' SEV-2016-0588, SEV-2015-0548 and SEV-2012-0234, the Unidad de Excelencia ``Mar\'{\i}a de Maeztu'' MDM-2014-0369 and the "la Caixa" Foundation (fellowship LCF/BQ/PI18/11630012), by the Croatian Science Foundation (HrZZ) Project IP-2016-06-9782 and the University of Rijeka Project 13.12.1.3.02, by the DFG Collaborative Research Centers SFB823/C4 and SFB876/C3, the Polish National Research Centre grant UMO-2016/22/M/ST9/00382 and by the Brazilian MCTIC, CNPq and FAPERJ.
K.~Noda is thankful to the support by Marie Sk\l odowska-Curie actions (H2020-MSCA-COFUND-2014, Project P-Sphere GA 665919).
L.~Nava acknowledges funding from the European Union's Horizon 2020 research and innovation programme under the Marie Sk\l odowska-Curie grant agreement n.664931.
S.~Inoue is supported by JSPS KAKENHI grant number JP17K05460 from MEXT, Japan, and the RIKEN iTHEMS program. 

The \textit{Fermi} LAT Collaboration acknowledges generous ongoing support
from a number of agencies and institutes that have supported both the
development and the operation of the LAT as well as scientific data analysis.
These include the National Aeronautics and Space Administration and the
Department of Energy in the United States, the Commissariat \`a l'Energie Atomique
and the Centre National de la Recherche Scientifique / Institut National de Physique
Nucl\'eaire et de Physique des Particules in France, the Agenzia Spaziale Italiana
and the Istituto Nazionale di Fisica Nucleare in Italy, the Ministry of Education,
Culture, Sports, Science and Technology (MEXT), High Energy Accelerator Research
Organization (KEK) and Japan Aerospace Exploration Agency (JAXA) in Japan, and
the K.~A.~Wallenberg Foundation, the Swedish Research Council and the
Swedish National Space Board in Sweden.
 
Additional support for science analysis during the operations phase is gratefully 
acknowledged from the Istituto Nazionale di Astrofisica in Italy and the Centre 
National d'\'Etudes Spatiales in France. This work performed in part under DOE 
Contract DE-AC02-76SF00515.

\bibliography{GRB160821B-paper}

\begin{thebibliography}{}
\expandafter\ifx\csname natexlab\endcsname\relax\def\natexlab#1{#1}\fi
\providecommand{\url}[1]{\href{#1}{#1}}
\providecommand{\dodoi}[1]{doi:~\href{http://doi.org/#1}{\nolinkurl{#1}}}
\providecommand{\doeprint}[1]{\href{http://ascl.net/#1}{\nolinkurl{http://ascl.net/#1}}}
\providecommand{\doarXiv}[1]{\href{https://arxiv.org/abs/#1}{\nolinkurl{https://arxiv.org/abs/#1}}}

\bibitem[{{Abbott} {et~al.}(2017)}]{Abbott17a}
{Abbott}, B.~P., {et~al.} 2017, \apjl, 848, L12,
  \dodoi{10.3847/2041-8213/aa91c9}

\bibitem[{{Abdalla} {et~al.}(2019){Abdalla}, {Adam}, {Aharonian}, {Ait
  Benkhali}, {Ang{\"u}ner}, {Arakawa}, {Arcaro}, {Armand }, {Ashkar}, {Backes},
  {Barbosa Martins}, {Barnard}, {Becherini}, {Berge}, {Bernl{\"o}hr},
  {Bissaldi}, {Blackwell}, {B{\"o}ttcher}, {Boisson}, {Bolmont}, {Bonnefoy},
  {Bregeon}, {Breuhaus}, {Brun}, {Brun}, {Bryan}, {B{\"u}chele}, {Bulik},
  {Bylund}, {Capasso}, {Caroff}, {Carosi}, {Casanova}, {Cerruti}, {Chand},
  {Chandra}, {Chen}, {Colafrancesco}, {Cury{\l}o}, {Davids}, {Deil}, {Devin},
  {deWilt}, {Dirson}, {Djannati-Ata{\"\i}}, {Dmytriiev}, {Donath},
  {Doroshenko}, {Dyks}, {Egberts}, {Emery}, {Ernenwein}, {Eschbach}, {Feijen},
  {Fegan}, {Fiasson}, {Fontaine}, {Funk}, {F{\"u}{\ss}ling}, {Gabici},
  {Gallant}, {Gat{\'e}}, {Giavitto}, {Giunti}, {Glawion}, {Glicenstein},
  {Gottschall}, {Grondin}, {Hahn}, {Haupt}, {Heinzelmann}, {Henri}, {Hermann},
  {Hinton}, {Hofmann}, {Hoischen}, {Holch}, {Holler}, {Horns}, {Huber},
  {Iwasaki}, {Jamrozy}, {Jankowsky}, {Jankowsky}, {Jardin-Blicq},
  {Jung-Richardt}, {Kastendieck}, {Katarzy{\'n}ski}, {Katsuragawa}, {Katz},
  {Khangulyan}, {Kh{\'e}lifi}, {King}, {Klepser}, {Klu{\'z}niak}, {Komin},
  {Kosack}, {Kostunin}, {Kreter}, {Lamanna}, {Lemi{\`e}re}, {Lemoine-Goumard},
  {Lenain}, {Leser}, {Levy}, {Lohse}, {Lypova}, {Mackey}, {Majumdar},
  {Malyshev}, {Marandon}, {Marcowith}, {Mares}, {Mariaud}, {Mart{\'\i}-Devesa},
  {Marx}, {Maurin}, {Meintjes}, {Mitchell}, {Moderski}, {Mohamed}, {Mohrmann},
  {Moore}, {Moulin}, {Muller}, {Murach}, {Nakashima}, {de Naurois},
  {Ndiyavala}, {Niederwanger}, {Niemiec}, {Oakes}, {O'Brien}, {Odaka}, {Ohm},
  {de Ona Wilhelmi}, {Ostrowski}, {Oya}, {Panter}, {Parsons}, {Perennes},
  {Petrucci}, {Peyaud}, {Piel}, {Pita}, {Poireau}, {Priyana Noel}, {Prokhorov},
  {Prokoph}, {P{\"u}hlhofer}, {Punch}, {Quirrenbach}, {Raab}, {Rauth},
  {Reimer}, {Reimer}, {Remy}, {Renaud}, {Rieger}, {Rinchiuso}, {Romoli},
  {Rowell}, {Rudak}, {Ruiz-Velasco}, {Sahakian}, {Sailer}, {Saito}, {Sanchez},
  {Santangelo}, {Sasaki}, {Schlickeiser}, {Sch{\"u}ssler}, {Schulz}, {Schutte},
  {Schwanke}, {Schwemmer}, {Seglar-Arroyo}, {Senniappan}, {Seyffert}, {Shafi},
  {Shiningayamwe}, {Simoni}, {Sinha}, {Sol}, {Specovius}, {Spir-Jacob},
  {Stawarz}, {Steenkamp}, {Stegmann}, {Steppa}, {Takahashi}, {Tavernier},
  {Taylor}, {Terrier}, {Tiziani}, {Tluczykont}, {Trichard}, {Tsirou}, {Tsuji},
  {Tuffs}, {Uchiyama}, {van der Walt}, {van Eldik}, {van Rensburg}, {van
  Soelen}, {Vasileiadis}, {Veh}, {Venter}, {Vincent}, {Vink}, {V{\"o}lk},
  {Vuillaume}, {Wadiasingh}, {Wagner}, {White}, {Wierzcholska}, {Yang},
  {Yoneda}, {Zacharias}, {Zanin}, {Zdziarski}, {Zech}, {Ziegler}, {Zorn},
  {{\.Z}ywucka}, {de Palma}, {Axelsson}, \& {Roberts}}]{HESS19}
{Abdalla}, H., {Adam}, R., {Aharonian}, F., {et~al.} 2019, \nat, 575, 464,
  \dodoi{10.1038/s41586-019-1743-9}

\bibitem[{{Abdollahi} {et~al.}(2020){Abdollahi}, {Acero}, {Ackermann},
  {Ajello}, {Atwood}, {Axelsson}, {Baldini}, {Ballet}, {Barbiellini},
  {Bastieri}, {Becerra Gonzalez}, {Bellazzini}, {Berretta}, {Bissaldi}, {Bland
  ford}, {Bloom}, {Bonino}, {Bottacini}, {Brandt}, {Bregeon}, {Bruel},
  {Buehler}, {Burnett}, {Buson}, {Cameron}, {Caputo}, {Caraveo}, {Casandjian},
  {Castro}, {Cavazzuti}, {Charles}, {Chaty}, {Chen}, {Cheung}, {Chiaro},
  {Ciprini}, {Cohen-Tanugi}, {Cominsky}, {Coronado-Bl{\'a}zquez}, {Costantin},
  {Cuoco}, {Cutini}, {D'Ammando}, {DeKlotz}, {de la Torre Luque}, {de Palma},
  {Desai}, {Digel}, {Di Lalla}, {Di Mauro}, {Di Venere}, {Dom{\'\i}nguez},
  {Dumora}, {Fana Dirirsa}, {Fegan}, {Ferrara}, {Franckowiak}, {Fukazawa},
  {Funk}, {Fusco}, {Gargano}, {Gasparrini}, {Giglietto}, {Giommi}, {Giordano},
  {Giroletti}, {Glanzman}, {Green}, {Grenier}, {Griffin}, {Grondin}, {Grove},
  {Guiriec}, {Harding}, {Hayashi}, {Hays}, {Hewitt}, {Horan},
  {J{\'o}hannesson}, {Johnson}, {Kamae}, {Kerr}, {Kocevski}, {Kovac'evic'},
  {Kuss}, {Landriu}, {Larsson}, {Latronico}, {Lemoine-Goumard}, {Li},
  {Liodakis}, {Longo}, {Loparco}, {Lott}, {Lovellette}, {Lubrano}, {Madejski},
  {Maldera}, {Malyshev}, {Manfreda}, {Marchesini}, {Marcotulli},
  {Mart{\'\i}-Devesa}, {Martin}, {Massaro}, {Mazziotta}, {McEnery}, {Mereu},
  {Meyer}, {Michelson}, {Mirabal}, {Mizuno}, {Monzani}, {Morselli},
  {Moskalenko}, {Negro}, {Nuss}, {Ojha}, {Omodei}, {Orienti}, {Orlando},
  {Ormes}, {Palatiello}, {Paliya}, {Paneque}, {Pei}, {Pe{\~n}a-Herazo},
  {Perkins}, {Persic}, {Pesce-Rollins}, {Petrosian}, {Petrov}, {Piron}, {Poon},
  {Porter}, {Principe}, {Rain{\`o}}, {Rando}, {Razzano}, {Razzaque}, {Reimer},
  {Reimer}, {Remy}, {Reposeur}, {Romani}, {Saz Parkinson}, {Schinzel},
  {Serini}, {Sgr{\`o}}, {Siskind}, {Smith}, {Spandre}, {Spinelli}, {Strong},
  {Suson}, {Tajima}, {Takahashi}, {Tak}, {Thayer}, {Thompson}, {Tibaldo},
  {Torres}, {Torresi}, {Valverde}, {Van Klaveren}, {van Zyl}, {Wood},
  {Yassine}, \& {Zaharijas}}]{4FGL}
{Abdollahi}, S., {Acero}, F., {Ackermann}, M., {et~al.} 2020, \apjs, 247, 33,
  \dodoi{10.3847/1538-4365/ab6bcb}

\bibitem[{{Acciari} {et~al.}(2020){Acciari}, {Ansoldi}, {Antonelli}, {Arbet
  Engels}, {Baack}, {Babi{\'c}}, {Banerjee}, {Barres de Almeida}, {Barrio},
  {Becerra Gonz{\'a}lez}, {Bednarek}, {Bellizzi}, {Bernardini}, {Berti},
  {Besenrieder}, {Bhattacharyya}, {Bigongiari}, {Biland}, {Blanch}, {Bonnoli},
  {Bo{\v{s}}njak}, {Busetto}, {Carosi}, {Ceribella}, {Cerruti}, {Chai},
  {Chilingarian}, {Cikota}, {Colak}, {Colin}, {Colombo}, {Contreras},
  {Cortina}, {Covino}, {D'Amico}, {D'Elia}, {da Vela}, {Dazzi}, {de Angelis},
  {de Lotto}, {Delfino}, {Delgado}, {Depaoli}, {di Pierro}, {di Venere}, {Do
  Souto Espi{\~n}eira}, {Dominis Prester}, {Donini}, {Dorner}, {Doro},
  {Elsaesser}, {Fallah Ramazani}, {Fattorini}, {Ferrara}, {Foffano}, {Fonseca},
  {Font}, {Fruck}, {Fukami}, {Garc{\'\i}a L{\'o}pez}, {Garczarczyk},
  {Gasparyan}, {Gaug}, {Giglietto}, {Giordano}, {Gliwny}, {Godinovi{\'c}},
  {Green}, {Hadasch}, {Hahn}, {Herrera}, {Hoang}, {Hrupec}, {H{\"u}tten},
  {Inada}, {Inoue}, {Ishio}, {Iwamura}, {Jouvin}, {Kajiwara}, {Karjalainen},
  {Kerszberg}, {Kobayashi}, {Kubo}, {Kushida}, {Lamastra}, {Lelas}, {Leone},
  {Lindfors}, {Lombardi}, {Longo}, {L{\'o}pez}, {L{\'o}pez-Coto},
  {L{\'o}pez-Oramas}, {Loporchio}, {Machado de Oliveira Fraga}, {Maggio},
  {Majumdar}, {Makariev}, {Mallamaci}, {Maneva}, {Manganaro}, {Mannheim},
  {Maraschi}, {Mariotti}, {Mart{\'\i}nez}, {Mazin}, {Mender},
  {Mi{\'c}anovi{\'c}}, {Miceli}, {Miener}, {Minev}, {Miranda}, {Mirzoyan},
  {Molina}, {Moralejo}, {Morcuende}, {Moreno}, {Moretti}, {Munar-Adrover},
  {Neustroev}, {Nigro}, {Nilsson}, {Ninci}, {Nishijima}, {Noda}, {Nogu{\'e}s},
  {Nozaki}, {Ohtani}, {Oka}, {Otero-Santos}, {Palatiello}, {Paneque},
  {Paoletti}, {Paredes}, {Pavleti{\'c}}, {Pe{\~n}il}, {Perennes}, {Peresano},
  {Persic}, {Prada Moroni}, {Prand ini}, {Puljak}, {Rhode}, {Rib{\'o}}, {Rico},
  {Righi}, {Rugliancich}, {Saha}, {Sahakyan}, {Saito}, {Sakurai}, {Satalecka},
  {Schleicher}, {Schmidt}, {Schweizer}, {Sitarek}, {{\v{S}}nidari{\'c}},
  {Sobczynska}, {Spolon}, {Stamerra}, {Strom}, {Strzys}, {Suda}, {Suri{\'c}},
  {Takahashi}, {Tavecchio}, {Temnikov}, {Terzi{\'c}}, {Teshima},
  {Torres-Alb{\`a}}, {Tosti}, {van Scherpenberg}, {Vanzo}, {Vazquez Acosta},
  {Ventura}, {Verguilov}, {Vigorito}, {Vitale}, {Vovk}, {Will}, {Zari{\'c}},
  {Nava}, \& {MAGIC Collaboration}}]{MAGIC-190114C-c}
{Acciari}, V.~A., {Ansoldi}, S., {Antonelli}, L.~A., {et~al.} 2020, \prl, 125,
  021301, \dodoi{10.1103/PhysRevLett.125.021301}

\bibitem[{{Ackermann} {et~al.}(2010){Ackermann}, {Asano}, {Atwood}, {Axelsson},
  {Baldini}, {Ballet}, {Barbiellini}, {Baring}, {Bastieri}, {Bechtol},
  {Bellazzini}, {Berenji}, {Bhat}, {Bissaldi}, {Blandford}, {Bloom},
  {Bonamente}, {Borgland}, {Bouvier}, {Bregeon}, {Brez}, {Briggs}, {Brigida},
  {Bruel}, {Buson}, {Caliandro}, {Cameron}, {Caraveo}, {Carrigan}, {Casand
  jian}, {Cecchi}, {{\c{C}}elik}, {Charles}, {Chiang}, {Ciprini}, {Claus},
  {Cohen-Tanugi}, {Connaughton}, {Conrad}, {Dermer}, {de Palma}, {Dingus},
  {Silva}, {Drell}, {Dubois}, {Dumora}, {Farnier}, {Favuzzi}, {Fegan}, {Finke},
  {Focke}, {Frailis}, {Fukazawa}, {Fusco}, {Gargano}, {Gasparrini}, {Gehrels},
  {Germani}, {Giglietto}, {Giordano}, {Glanzman}, {Godfrey}, {Granot},
  {Grenier}, {Grondin}, {Grove}, {Guiriec}, {Hadasch}, {Harding}, {Hays},
  {Horan}, {Hughes}, {J{\'o}hannesson}, {Johnson}, {Kamae}, {Katagiri},
  {Kataoka}, {Kawai}, {Kippen}, {Kn{\"o}dlseder}, {Kocevski}, {Kouveliotou},
  {Kuss}, {Lande}, {Latronico}, {Lemoine-Goumard}, {Llena Garde}, {Longo},
  {Loparco}, {Lott}, {Lovellette}, {Lubrano}, {Makeev}, {Mazziotta}, {McEnery},
  {McGlynn}, {Meegan}, {M{\'e}sz{\'a}ros}, {Michelson}, {Mitthumsiri},
  {Mizuno}, {Moiseev}, {Monte}, {Monzani}, {Moretti}, {Morselli}, {Moskalenko},
  {Murgia}, {Nakajima}, {Nakamori}, {Nolan}, {Norris}, {Nuss}, {Ohno},
  {Ohsugi}, {Omodei}, {Orlando}, {Ormes}, {Ozaki}, {Paciesas}, {Paneque},
  {Panetta}, {Parent}, {Pelassa}, {Pepe}, {Pesce-Rollins}, {Piron}, {Preece},
  {Rain{\`o}}, {Rando}, {Razzano}, {Razzaque}, {Reimer}, {Ritz}, {Rodriguez},
  {Roth}, {Ryde}, {Sadrozinski}, {Sander}, {Scargle}, {Schalk}, {Sgr{\`o}},
  {Siskind}, {Smith}, {Spandre}, {Spinelli}, {Stamatikos}, {Stecker},
  {Strickman}, {Suson}, {Tajima}, {Takahashi}, {Takahashi}, {Tanaka}, {Thayer},
  {Thayer}, {Thompson}, {Tibaldo}, {Toma}, {Torres}, {Tosti}, {Tramacere},
  {Uchiyama}, {Uehara}, {Usher}, {van der Horst}, {Vasileiou}, {Vilchez},
  {Vitale}, {von Kienlin}, {Waite}, {Wang}, {Wilson-Hodge}, {Winer}, {Wu},
  {Yamazaki}, {Yang}, {Ylinen}, \& {Ziegler}}]{Ackermann10}
{Ackermann}, M., {Asano}, K., {Atwood}, W.~B., {et~al.} 2010, \apj, 716, 1178,
  \dodoi{10.1088/0004-637X/716/2/1178}

\bibitem[{{Ahnen} {et~al.}(2017){Ahnen}, {Ansoldi}, {Antonelli}, {Arcaro},
  {Babi{\'c}}, {Banerjee}, {Bangale}, {Barres de Almeida}, {Barrio}, {Becerra
  Gonz{\'a}lez}, {Bednarek}, {Bernardini}, {Berti}, {Bhattacharyya},
  {Biasuzzi}, {Biland }, {Blanch}, {Bonnefoy}, {Bonnoli}, {Carosi}, {Carosi},
  {Chatterjee}, {Colin}, {Colombo}, {Contreras}, {Cortina}, {Covino}, {Cumani},
  {Da Vela}, {Dazzi}, {De Angelis}, {De Lotto}, {de O{\~n}a Wilhelmi}, {Di
  Pierro}, {Doert}, {Dom{\'\i}nguez}, {Dominis Prester}, {Dorner}, {Doro},
  {Einecke}, {Eisenacher Glawion}, {Elsaesser}, {Engelkemeier}, {Fallah
  Ramazani}, {Fern{\'a}ndez-Barral}, {Fidalgo}, {Fonseca}, {Font}, {Fruck},
  {Galindo}, {Garc{\'\i}a L{\'o}pez}, {Garczarczyk}, {Gaug}, {Giammaria},
  {Godinovi{\'c}}, {Gora}, {Griffiths}, {Guberman}, {Hadasch}, {Hahn},
  {Hassan}, {Hayashida}, {Herrera}, {Hose}, {Hrupec}, {Hughes}, {Ishio},
  {Konno}, {Kubo}, {Kushida}, {Kuve{\v{z}}di{\'c}}, {Lelas}, {Lindfors},
  {Lombardi}, {Longo}, {L{\'o}pez}, {Maggio}, {Majumdar}, {Makariev}, {Maneva},
  {Manganaro}, {Mannheim}, {Maraschi}, {Mariotti}, {Mart{\'\i}nez}, {Mazin},
  {Menzel}, {Minev}, {Mirzoyan}, {Moralejo}, {Moreno}, {Moretti}, {Neustroev},
  {Niedzwiecki}, {Nievas Rosillo}, {Nilsson}, {Ninci}, {Nishijima}, {Noda},
  {Nogu{\'e}s}, {Paiano}, {Palacio}, {Paneque}, {Paoletti}, {Paredes},
  {Paredes-Fortuny}, {Pedaletti}, {Peresano}, {Perri}, {Persic}, {Prada
  Moroni}, {Prandini}, {Puljak}, {Garcia}, {Reichardt}, {Rhode}, {Rib{\'o}},
  {Rico}, {Rugliancich}, {Saito}, {Satalecka}, {Schroeder}, {Schweizer},
  {Sillanp{\"a}{\"a}}, {Sitarek}, {{\v{S}}nidari{\'c}}, {Sobczynska},
  {Stamerra}, {Strzys}, {Suri{\'c}}, {Takalo}, {Tavecchio}, {Temnikov},
  {Terzi{\'c}}, {Tescaro}, {Teshima}, {Torres}, {Torres-Alb{\`a}}, {Treves},
  {Vanzo}, {Vazquez Acosta}, {Vovk}, {Ward}, {Will}, \&
  {Zari{\'c}}}]{MoonPerformance}
{Ahnen}, M.~L., {Ansoldi}, S., {Antonelli}, L.~A., {et~al.} 2017, Astroparticle
  Physics, 94, 29, \dodoi{10.1016/j.astropartphys.2017.08.001}

\bibitem[{{Ajello} {et~al.}(2019{\natexlab{a}}){Ajello}, {Arimoto}, {Axelsson},
  {Baldini}, {Barbiellini}, {Bastieri}, {Bellazzini}, {Bhat}, {Bissaldi},
  {Blandford}, {Bonino}, {Bonnell}, {Bottacini}, {Bregeon}, {Bruel}, {Buehler},
  {Cameron}, {Caputo}, {Caraveo}, {Cavazzuti}, {Chen}, {Cheung}, {Chiaro},
  {Ciprini}, {Costantin}, {Crnogorcevic}, {Cutini}, {Dainotti}, {D'Ammand o},
  {de la Torre Luque}, {de Palma}, {Desai}, {Desiante}, {Di Lalla}, {Di
  Venere}, {Fana Dirirsa}, {Fegan}, {Franckowiak}, {Fukazawa}, {Funk}, {Fusco},
  {Gargano}, {Gasparrini}, {Giglietto}, {Giordano}, {Giroletti}, {Green},
  {Grenier}, {Grove}, {Guiriec}, {Hays}, {Hewitt}, {Horan}, {J{\'o}hannesson},
  {Kocevski}, {Kuss}, {Latronico}, {Li}, {Longo}, {Loparco}, {Lovellette},
  {Lubrano}, {Maldera}, {Manfreda}, {Mart{\'\i}-Devesa}, {Mazziotta}, {Mereu},
  {Meyer}, {Michelson}, {Mirabal}, {Mitthumsiri}, {Mizuno}, {Monzani},
  {Moretti}, {Morselli}, {Moskalenko}, {Negro}, {Nuss}, {Ohno}, {Omodei},
  {Orienti}, {Orlando}, {Palatiello}, {Paliya}, {Paneque}, {Persic},
  {Pesce-Rollins}, {Petrosian}, {Piron}, {Poolakkil}, {Poon}, {Porter},
  {Principe}, {Racusin}, {Rain{\`o}}, {Rando}, {Razzano}, {Razzaque}, {Reimer},
  {Reimer}, {Reposeur}, {Ryde}, {Serini}, {Sgr{\`o}}, {Siskind}, {Sonbas},
  {Spandre}, {Spinelli}, {Suson}, {Tajima}, {Takahashi}, {Tak}, {Thayer},
  {Torres}, {Troja}, {Valverde}, {Veres}, {Vianello}, {von Kienlin}, {Wood},
  {Yassine}, {Zhu}, \& {Zimmer}}]{Ajello19}
{Ajello}, M., {Arimoto}, M., {Axelsson}, M., {et~al.} 2019{\natexlab{a}}, \apj,
  878, 52, \dodoi{10.3847/1538-4357/ab1d4e}

\bibitem[{{Ajello} {et~al.}(2019{\natexlab{b}}){Ajello}, {Arimoto}, {Axelsson},
  {Baldini}, {Barbiellini}, {Bastieri}, {Bellazzini}, {Bhat}, {Bissaldi},
  {Blandford}, {Bonino}, {Bonnell}, {Bottacini}, {Bregeon}, {Bruel}, {Buehler},
  {Cameron}, {Caputo}, {Caraveo}, {Cavazzuti}, {Chen}, {Cheung}, {Chiaro},
  {Ciprini}, {Costantin}, {Crnogorcevic}, {Cutini}, {Dainotti}, {D'Ammand o},
  {de la Torre Luque}, {de Palma}, {Desai}, {Desiante}, {Di Lalla}, {Di
  Venere}, {Fana Dirirsa}, {Fegan}, {Franckowiak}, {Fukazawa}, {Funk}, {Fusco},
  {Gargano}, {Gasparrini}, {Giglietto}, {Giordano}, {Giroletti}, {Green},
  {Grenier}, {Grove}, {Guiriec}, {Hays}, {Hewitt}, {Horan}, {J{\'o}hannesson},
  {Kocevski}, {Kuss}, {Latronico}, {Li}, {Longo}, {Loparco}, {Lovellette},
  {Lubrano}, {Maldera}, {Manfreda}, {Mart{\'\i}-Devesa}, {Mazziotta}, {Mereu},
  {Meyer}, {Michelson}, {Mirabal}, {Mitthumsiri}, {Mizuno}, {Monzani},
  {Moretti}, {Morselli}, {Moskalenko}, {Negro}, {Nuss}, {Ohno}, {Omodei},
  {Orienti}, {Orlando}, {Palatiello}, {Paliya}, {Paneque}, {Persic},
  {Pesce-Rollins}, {Petrosian}, {Piron}, {Poolakkil}, {Poon}, {Porter},
  {Principe}, {Racusin}, {Rain{\`o}}, {Rando}, {Razzano}, {Razzaque}, {Reimer},
  {Reimer}, {Reposeur}, {Ryde}, {Serini}, {Sgr{\`o}}, {Siskind}, {Sonbas},
  {Spandre}, {Spinelli}, {Suson}, {Tajima}, {Takahashi}, {Tak}, {Thayer},
  {Torres}, {Troja}, {Valverde}, {Veres}, {Vianello}, {von Kienlin}, {Wood},
  {Yassine}, {Zhu}, \& {Zimmer}}]{secondLATGRB}
---. 2019{\natexlab{b}}, \apj, 878, 52, \dodoi{10.3847/1538-4357/ab1d4e}

\bibitem[{{Aleksi{\'c}} {et~al.}(2016){Aleksi{\'c}}, {Ansoldi}, {Antonelli},
  {Antoranz}, {Babic}, {Bangale}, {Barcel{\'o}}, {Barrio}, {Becerra
  Gonz{\'a}lez}, {Bednarek}, {Bernardini}, {Biasuzzi}, {Biland}, {Bitossi},
  {Blanch}, {Bonnefoy}, {Bonnoli}, {Borracci}, {Bretz}, {Carmona}, {Carosi},
  {Cecchi}, {Colin}, {Colombo}, {Contreras}, {Corti}, {Cortina}, {Covino}, {Da
  Vela}, {Dazzi}, {De Angelis}, {De Caneva}, {De Lotto}, {de O{\~n}a Wilhelmi},
  {Delgado Mendez}, {Dettlaff}, {Dominis Prester}, {Dorner}, {Doro}, {Einecke},
  {Eisenacher}, {Elsaesser}, {Fidalgo}, {Fink}, {Fonseca}, {Font}, {Frantzen},
  {Fruck}, {Galindo}, {Garc{\'\i}a L{\'o}pez}, {Garczarczyk}, {Garrido
  Terrats}, {Gaug}, {Giavitto}, {Godinovi{\'c}}, {Gonz{\'a}lez Mu{\~n}oz},
  {Gozzini}, {Haberer}, {Hadasch}, {Hanabata}, {Hayashida}, {Herrera},
  {Hildebrand }, {Hose}, {Hrupec}, {Idec}, {Illa}, {Kadenius}, {Kellermann},
  {Knoetig}, {Kodani}, {Konno}, {Krause}, {Kubo}, {Kushida}, {La Barbera},
  {Lelas}, {Lemus}, {Lewandowska}, {Lindfors}, {Lombardi}, {Longo},
  {L{\'o}pez}, {L{\'o}pez-Coto}, {L{\'o}pez-Oramas}, {Lorca}, {Lorenz},
  {Lozano}, {Makariev}, {Mallot}, {Maneva}, {Mankuzhiyil}, {Mannheim},
  {Maraschi}, {Marcote}, {Mariotti}, {Mart{\'\i}nez}, {Mazin}, {Menzel},
  {Miranda}, {Mirzoyan}, {Moralejo}, {Munar-Adrover}, {Nakajima}, {Negrello},
  {Neustroev}, {Niedzwiecki}, {Nilsson}, {Nishijima}, {Noda}, {Orito},
  {Overkemping}, {Paiano}, {Palatiello}, {Paneque}, {Paoletti}, {Paredes},
  {Paredes-Fortuny}, {Persic}, {Poutanen}, {Prada Moroni}, {Prandini},
  {Puljak}, {Reinthal}, {Rhode}, {Rib{\'o}}, {Rico}, {Rodriguez Garcia},
  {R{\"u}gamer}, {Saito}, {Saito}, {Satalecka}, {Scalzotto}, {Scapin},
  {Schultz}, {Schlammer}, {Schmidl}, {Schweizer}, {Shore}, {Sillanp{\"a}{\"a}},
  {Sitarek}, {Snidaric}, {Sobczynska}, {Spanier}, {Stamerra}, {Steinbring},
  {Storz}, {Strzys}, {Takalo}, {Takami}, {Tavecchio}, {Tejedor}, {Temnikov},
  {Terzi{\'c}}, {Tescaro}, {Teshima}, {Thaele}, {Tibolla}, {Torres}, {Toyama},
  {Treves}, {Vogler}, {Wetteskind}, {Will}, \& {Zanin}}]{PerformancePaperII}
{Aleksi{\'c}}, J., {Ansoldi}, S., {Antonelli}, L.~A., {et~al.} 2016,
  Astroparticle Physics, 72, 76, \dodoi{10.1016/j.astropartphys.2015.02.005}

\bibitem[{{Atwood} {et~al.}(2013){Atwood}, {Albert}, {Baldini}, {Tinivella},
  {Bregeon}, {Pesce-Rollins}, {Sgr{\`o}}, {Bruel}, {Charles}, {Drlica-Wagner},
  {Franckowiak}, {Jogler}, {Rochester}, {Usher}, {Wood}, {Cohen-Tanugi}, \&
  {Zimmer}}]{pass8}
{Atwood}, W., {Albert}, A., {Baldini}, L., {et~al.} 2013, arXiv e-prints,
  arXiv:1303.3514.
\newblock \doarXiv{1303.3514}

\bibitem[{{Atwood} {et~al.}(2009){Atwood}, {Abdo}, {Ackermann}, {Althouse},
  {Anderson}, {Axelsson}, {Baldini}, {Ballet}, {Band}, {Barbiellini},
  {Bartelt}, {Bastieri}, {Baughman}, {Bechtol}, {B{\'e}d{\'e}r{\`e}de},
  {Bellardi}, {Bellazzini}, {Berenji}, {Bignami}, {Bisello}, {Bissaldi},
  {Blandford}, {Bloom}, {Bogart}, {Bonamente}, {Bonnell}, {Borgland },
  {Bouvier}, {Bregeon}, {Brez}, {Brigida}, {Bruel}, {Burnett}, {Busetto},
  {Caliandro}, {Cameron}, {Caraveo}, {Carius}, {Carlson}, {Casandjian},
  {Cavazzuti}, {Ceccanti}, {Cecchi}, {Charles}, {Chekhtman}, {Cheung},
  {Chiang}, {Chipaux}, {Cillis}, {Ciprini}, {Claus}, {Cohen-Tanugi},
  {Condamoor}, {Conrad}, {Corbet}, {Corucci}, {Costamante}, {Cutini}, {Davis},
  {Decotigny}, {DeKlotz}, {Dermer}, {de Angelis}, {Digel}, {do Couto e Silva},
  {Drell}, {Dubois}, {Dumora}, {Edmonds}, {Fabiani}, {Farnier}, {Favuzzi},
  {Flath}, {Fleury}, {Focke}, {Funk}, {Fusco}, {Gargano}, {Gasparrini},
  {Gehrels}, {Gentit}, {Germani}, {Giebels}, {Giglietto}, {Giommi}, {Giordano},
  {Glanzman}, {Godfrey}, {Grenier}, {Grondin}, {Grove}, {Guillemot}, {Guiriec},
  {Haller}, {Harding}, {Hart}, {Hays}, {Healey}, {Hirayama}, {Hjalmarsdotter},
  {Horn}, {Hughes}, {J{\'o}hannesson}, {Johansson}, {Johnson}, {Johnson},
  {Johnson}, {Johnson}, {Kamae}, {Katagiri}, {Kataoka}, {Kavelaars}, {Kawai},
  {Kelly}, {Kerr}, {Klamra}, {Kn{\"o}dlseder}, {Kocian}, {Komin}, {Kuehn},
  {Kuss}, {Landriu}, {Latronico}, {Lee}, {Lee}, {Lemoine-Goumard}, {Lionetto},
  {Longo}, {Loparco}, {Lott}, {Lovellette}, {Lubrano}, {Madejski}, {Makeev},
  {Marangelli}, {Massai}, {Mazziotta}, {McEnery}, {Menon}, {Meurer},
  {Michelson}, {Minuti}, {Mirizzi}, {Mitthumsiri}, {Mizuno}, {Moiseev},
  {Monte}, {Monzani}, {Moretti}, {Morselli}, {Moskalenko}, {Murgia},
  {Nakamori}, {Nishino}, {Nolan}, {Norris}, {Nuss}, {Ohno}, {Ohsugi}, {Omodei},
  {Orlando}, {Ormes}, {Paccagnella}, {Paneque}, {Panetta}, {Parent}, {Pearce},
  {Pepe}, {Perazzo}, {Pesce-Rollins}, {Picozza}, {Pieri}, {Pinchera}, {Piron},
  {Porter}, {Poupard}, {Rain{\`o}}, {Rando}, {Rapposelli}, {Razzano}, {Reimer},
  {Reimer}, {Reposeur}, {Reyes}, {Ritz}, {Rochester}, {Rodriguez}, {Romani},
  {Roth}, {Russell}, {Ryde}, {Sabatini}, {Sadrozinski}, {Sanchez}, {Sand er},
  {Sapozhnikov}, {Parkinson}, {Scargle}, {Schalk}, {Scolieri}, {Sgr{\`o}},
  {Share}, {Shaw}, {Shimokawabe}, {Shrader}, {Sierpowska-Bartosik}, {Siskind},
  {Smith}, {Smith}, {Spandre}, {Spinelli}, {Starck}, {Stephens}, {Strickman},
  {Strong}, {Suson}, {Tajima}, {Takahashi}, {Takahashi}, {Tanaka}, {Tenze},
  {Tether}, {Thayer}, {Thayer}, {Thompson}, {Tibaldo}, {Tibolla}, {Torres},
  {Tosti}, {Tramacere}, {Turri}, {Usher}, {Vilchez}, {Vitale}, {Wang},
  {Watters}, {Winer}, {Wood}, {Ylinen}, \& {Ziegler}}]{LAT-instrument}
{Atwood}, W.~B., {Abdo}, A.~A., {Ackermann}, M., {et~al.} 2009, \apj, 697,
  1071, \dodoi{10.1088/0004-637X/697/2/1071}

\bibitem[{{Barthelmy} {et~al.}(2005){Barthelmy}, {Barbier}, {Cummings},
  {Fenimore}, {Gehrels}, {Hullinger}, {Krimm}, {Markwardt}, {Palmer},
  {Parsons}, {Sato}, {Suzuki}, {Takahashi}, {Tashiro}, \& {Tueller}}]{BAT}
{Barthelmy}, S.~D., {Barbier}, L.~M., {Cummings}, J.~R., {et~al.} 2005, \ssr,
  120, 143, \dodoi{10.1007/s11214-005-5096-3}

\bibitem[{{Berger}(2014)}]{Berger14}
{Berger}, E. 2014, \araa, 52, 43, \dodoi{10.1146/annurev-astro-081913-035926}

\bibitem[{{Berti} {et~al.}(2019){Berti}, {Antonelli}, {Bosnjak}, {Cortina},
  {Covino}, {D'Elia}, {Espi{\~n}eira}, {Fukami}, {Inoue}, {Longo}, {Miceli},
  {Moretti}, {Nava}, {Noda}, {Nozaki}, {Peresano}, {Suda}, \& {Will}}]{Berti19}
{Berti}, A., {Antonelli}, L.~A., {Bosnjak}, Z., {et~al.} 2019, in International
  Cosmic Ray Conference, Vol.~36, 36th International Cosmic Ray Conference
  (ICRC2019), 634

\bibitem[{{B{\"o}ttcher} \& {Dermer}(1998)}]{Boettcher98}
{B{\"o}ttcher}, M., \& {Dermer}, C.~D. 1998, \apjl, 499, L131,
  \dodoi{10.1086/311366}

\bibitem[{{Burrows} {et~al.}(2005){Burrows}, {Hill}, {Nousek}, {Kennea},
  {Wells}, {Osborne}, {Abbey}, {Beardmore}, {Mukerjee}, {Short}, {Chincarini},
  {Campana}, {Citterio}, {Moretti}, {Pagani}, {Tagliaferri}, {Giommi},
  {Capalbi}, {Tamburelli}, {Angelini}, {Cusumano}, {Br{\"a}uninger}, {Burkert},
  \& {Hartner}}]{XRT}
{Burrows}, D.~N., {Hill}, J.~E., {Nousek}, J.~A., {et~al.} 2005, \ssr, 120,
  165, \dodoi{10.1007/s11214-005-5097-2}

\bibitem[{Carosi {et~al.}(2015)}]{MAGIC-GRB_ICRC15}
Carosi, A., {et~al.} 2015, Proc. of 34th ICRC 809

\bibitem[{{de Naurois} \& {H.~E.~S.~S. Collaboration}(2019)}]{HESS19-2}
{de Naurois}, M., \& {H.~E.~S.~S. Collaboration}. 2019, GRB Coordinates
  Network, 25566, 1

\bibitem[{{Dom{\'\i}nguez} {et~al.}(2011){Dom{\'\i}nguez}, {Primack},
  {Rosario}, {Prada}, {Gilmore}, {Faber}, {Koo}, {Somerville},
  {P{\'e}rez-Torres}, {P{\'e}rez-Gonz{\'a}lez}, {Huang}, {Davis},
  {Guhathakurta}, {Barmby}, {Conselice}, {Lozano}, {Newman}, \&
  {Cooper}}]{Dominguez11}
{Dom{\'\i}nguez}, A., {Primack}, J.~R., {Rosario}, D.~J., {et~al.} 2011,
  \mnras, 410, 2556, \dodoi{10.1111/j.1365-2966.2010.17631.x}

\bibitem[{{Dwek} \& {Krennrich}(2013)}]{Dwek13}
{Dwek}, E., \& {Krennrich}, F. 2013, Astroparticle Physics, 43, 112,
  \dodoi{10.1016/j.astropartphys.2012.09.003}

\bibitem[{{Evans} {et~al.}(2009){Evans}, {Beardmore}, {Page}, {Osborne},
  {O'Brien}, {Willingale}, {Starling}, {Burrows}, {Godet}, {Vetere}, {Racusin},
  {Goad}, {Wiersema}, {Angelini}, {Capalbi}, {Chincarini}, {Gehrels}, {Kennea},
  {Margutti}, {Morris}, {Mountford}, {Pagani}, {Perri}, {Romano}, \&
  {Tanvir}}]{Evans09}
{Evans}, P.~A., {Beardmore}, A.~P., {Page}, K.~L., {et~al.} 2009, \mnras, 397,
  1177, \dodoi{10.1111/j.1365-2966.2009.14913.x}

\bibitem[{Fong {et~al.}(2016)Fong, Alexander, \& Laskar}]{VLA-160821B}
Fong, W., Alexander, K.~D., \& Laskar, T. 2016, GCN Circ., 19854

\bibitem[{Fruck \& Gaug(2015)}]{Atmohead14}
Fruck, C., \& Gaug, M. 2015, Proc. of AtmoHEAD 2014, 02003

\bibitem[{{Gehrels} {et~al.}(2004){Gehrels}, {Chincarini}, {Giommi}, {Mason},
  {Nousek}, {Wells}, {White}, {Barthelmy}, {Burrows}, {Cominsky}, {Hurley},
  {Marshall}, {M{\'e}sz{\'a}ros}, {Roming}, {Angelini}, {Barbier}, {Belloni},
  {Campana}, {Caraveo}, {Chester}, {Citterio}, {Cline}, {Cropper}, {Cummings},
  {Dean}, {Feigelson}, {Fenimore}, {Frail}, {Fruchter}, {Garmire}, {Gendreau},
  {Ghisellini}, {Greiner}, {Hill}, {Hunsberger}, {Krimm}, {Kulkarni}, {Kumar},
  {Lebrun}, {Lloyd-Ronning}, {Markwardt}, {Mattson}, {Mushotzky}, {Norris},
  {Osborne}, {Paczynski}, {Palmer}, {Park}, {Parsons}, {Paul}, {Rees},
  {Reynolds}, {Rhoads}, {Sasseen}, {Schaefer}, {Short}, {Smale}, {Smith},
  {Stella}, {Tagliaferri}, {Takahashi}, {Tashiro}, {Townsley}, {Tueller},
  {Turner}, {Vietri}, {Voges}, {Ward}, {Willingale}, {Zerbi}, \&
  {Zhang}}]{Swift}
{Gehrels}, N., {Chincarini}, G., {Giommi}, P., {et~al.} 2004, \apj, 611, 1005,
  \dodoi{10.1086/422091}

\bibitem[{{Ghirlanda} {et~al.}(2019){Ghirlanda}, {Salafia}, {Paragi},
  {Giroletti}, {Yang}, {Marcote}, {Blanchard}, {Agudo}, {An}, {Bernardini},
  {Beswick}, {Branchesi}, {Campana}, {Casadio}, {Chassand e-Mottin}, {Colpi},
  {Covino}, {D'Avanzo}, {D'Elia}, {Frey}, {Gawronski}, {Ghisellini}, {Gurvits},
  {Jonker}, {van Langevelde}, {Melandri}, {Moldon}, {Nava}, {Perego},
  {Perez-Torres}, {Reynolds}, {Salvaterra}, {Tagliaferri}, {Venturi},
  {Vergani}, \& {Zhang}}]{Ghirlanda19}
{Ghirlanda}, G., {Salafia}, O.~S., {Paragi}, Z., {et~al.} 2019, Science, 363,
  968, \dodoi{10.1126/science.aau8815}

\bibitem[{{Ghisellini} {et~al.}(2005){Ghisellini}, {Tavecchio}, \&
  {Chiaberge}}]{Ghisellini05}
{Ghisellini}, G., {Tavecchio}, F., \& {Chiaberge}, M. 2005, \aap, 432, 401,
  \dodoi{10.1051/0004-6361:20041404}

\bibitem[{{Granot} \& {Sari}(2002)}]{granotsari}
{Granot}, J., \& {Sari}, R. 2002, \apj, 568, 820, \dodoi{10.1086/338966}

\bibitem[{Jeong {et~al.}(2016)}]{GTC-160821B}
Jeong, S., {et~al.} 2016, GCN Circ., 19847

\bibitem[{{Jin} {et~al.}(2020){Jin}, {Covino}, {Liao}, {Li}, {D'Avanzo}, {Fan},
  \& {Wei}}]{Jin20}
{Jin}, Z.-P., {Covino}, S., {Liao}, N.-H., {et~al.} 2020, Nature Astronomy, 4,
  77, \dodoi{10.1038/s41550-019-0892-y}

\bibitem[{{Jin} {et~al.}(2018){Jin}, {Li}, {Wang}, {Wang}, {He}, {Yuan},
  {Zhang}, {Zou}, {Fan}, \& {Wei}}]{Jin18}
{Jin}, Z.-P., {Li}, X., {Wang}, H., {et~al.} 2018, \apj, 857, 128,
  \dodoi{10.3847/1538-4357/aab76d}

\bibitem[{{Kasliwal} {et~al.}(2017){Kasliwal}, {Korobkin}, {Lau}, {Wollaeger},
  \& {Fryer}}]{Kasliwal17}
{Kasliwal}, M.~M., {Korobkin}, O., {Lau}, R.~M., {Wollaeger}, R., \& {Fryer},
  C.~L. 2017, \apjl, 843, L34, \dodoi{10.3847/2041-8213/aa799d}

\bibitem[{{Kimura} {et~al.}(2019){Kimura}, {Murase}, {Ioka}, {Kisaka}, {Fang},
  \& {M{\'e}sz{\'a}ros}}]{Kimura19}
{Kimura}, S.~S., {Murase}, K., {Ioka}, K., {et~al.} 2019, \apjl, 887, L16,
  \dodoi{10.3847/2041-8213/ab59e1}

\bibitem[{{Kumar} \& {Zhang}(2015)}]{Kumar15}
{Kumar}, P., \& {Zhang}, B. 2015, \physrep, 561, 1,
  \dodoi{10.1016/j.physrep.2014.09.008}

\bibitem[{{Lamb}(2020)}]{Lamb20}
{Lamb}, G.~P. 2020, arXiv e-prints, arXiv:2006.05893.
\newblock \doarXiv{2006.05893}

\bibitem[{{Lamb} {et~al.}(2019){Lamb}, {Tanvir}, {Levan}, {de Ugarte Postigo},
  {Kawaguchi}, {Corsi}, {Evans}, {Gompertz}, {Malesani}, {Page}, {Wiersema},
  {Rosswog}, {Shibata}, {Tanaka}, {van der Horst}, {Cano}, {Fynbo}, {Fruchter},
  {Greiner}, {Heintz}, {Higgins}, {Hjorth}, {Izzo}, {Jakobsson}, {Kann},
  {O'Brien}, {Perley}, {Pian}, {Pugliese}, {Starling}, {Th{\"o}ne}, {Watson},
  {Wijers}, \& {Xu}}]{Lamb19}
{Lamb}, G.~P., {Tanvir}, N.~R., {Levan}, A.~J., {et~al.} 2019, \apj, 883, 48,
  \dodoi{10.3847/1538-4357/ab38bb}

\bibitem[{Levan {et~al.}(2016)}]{WHT-160821B}
Levan, A.~J., {et~al.} 2016, GCN Circ., 19846

\bibitem[{{Li} \& {Ma}(1983)}]{LiMa}
{Li}, T.~P., \& {Ma}, Y.~Q. 1983, \apj, 272, 317, \dodoi{10.1086/161295}

\bibitem[{{Linial} \& {Sari}(2019)}]{Linial19}
{Linial}, I., \& {Sari}, R. 2019, \mnras, 483, 624,
  \dodoi{10.1093/mnras/sty3170}

\bibitem[{{Longo} {et~al.}(2016{\natexlab{a}}){Longo}, {Bissaldi}, {Bregeon},
  {McEnery}, {Ohno}, \& {Zhu}}]{2016GCN.19403....1L}
{Longo}, F., {Bissaldi}, E., {Bregeon}, J., {et~al.} 2016{\natexlab{a}}, GRB
  Coordinates Network, 19403, 1

\bibitem[{{Longo} {et~al.}(2016{\natexlab{b}}){Longo}, {Bissaldi}, {Vianello},
  {Moretti}, {Omodei}, {Bregeon}, {Dirirsa}, {Yassine}, {Kocevski}, {Racusin},
  {McEnery}, \& {Ohno}}]{2016GCN.19413....1L}
{Longo}, F., {Bissaldi}, E., {Vianello}, G., {et~al.} 2016{\natexlab{b}}, GRB
  Coordinates Network, 19413, 1

\bibitem[{{L{\"u}} {et~al.}(2015){L{\"u}}, {Zhang}, {Lei}, {Li}, \&
  {Lasky}}]{Lue15}
{L{\"u}}, H.-J., {Zhang}, B., {Lei}, W.-H., {Li}, Y., \& {Lasky}, P.~D. 2015,
  \apj, 805, 89, \dodoi{10.1088/0004-637X/805/2/89}

\bibitem[{{Madejski} \& {Sikora}(2016)}]{Madejski16}
{Madejski}, G.~G., \& {Sikora}, M. 2016, \araa, 54, 725,
  \dodoi{10.1146/annurev-astro-081913-040044}

\bibitem[{{MAGIC Collaboration} {et~al.}(2019{\natexlab{a}}){MAGIC
  Collaboration}, {Acciari}, {Ansoldi}, {Antonelli}, {Arbet Engels}, {Baack},
  {Babi{\'c}}, {Banerjee}, {Barres de Almeida}, {Barrio}, {Becerra
  Gonz{\'a}lez}, {Bednarek}, {Bellizzi}, {Bernardini}, {Berti}, {Besenrieder},
  {Bhattacharyya}, {Bigongiari}, {Biland }, {Blanch}, {Bonnoli},
  {Bo{\v{s}}njak}, {Busetto}, {Carosi}, {Carosi}, {Ceribella}, {Chai},
  {Chilingaryan}, {Cikota}, {Colak}, {Colin}, {Colombo}, {Contreras},
  {Cortina}, {Covino}, {D'Amico}, {D'Elia}, {da Vela}, {Dazzi}, {de Angelis},
  {de Lotto}, {Delfino}, {Delgado}, {Depaoli}, {di Pierro}, {di Venere}, {Do
  Souto Espi{\~n}eira}, {Dominis Prester}, {Donini}, {Dorner}, {Doro},
  {Elsaesser}, {Fallah Ramazani}, {Fattorini}, {Fern{\'a}ndez-Barral},
  {Ferrara}, {Fidalgo}, {Foffano}, {Fonseca}, {Font}, {Fruck}, {Fukami},
  {Gallozzi}, {Garc{\'\i}a L{\'o}pez}, {Garczarczyk}, {Gasparyan}, {Gaug},
  {Giglietto}, {Giordano}, {Godinovi{\'c}}, {Green}, {Guberman}, {Hadasch},
  {Hahn}, {Herrera}, {Hoang}, {Hrupec}, {H{\"u}tten}, {Inada}, {Inoue},
  {Ishio}, {Iwamura}, {Jouvin}, {Kerszberg}, {Kubo}, {Kushida}, {Lamastra},
  {Lelas}, {Leone}, {Lindfors}, {Lombardi}, {Longo}, {L{\'o}pez},
  {L{\'o}pez-Coto}, {L{\'o}pez-Oramas}, {Loporchio}, {Machado de Oliveira
  Fraga}, {Maggio}, {Majumdar}, {Makariev}, {Mallamaci}, {Maneva}, {Manganaro},
  {Mannheim}, {Maraschi}, {Mariotti}, {Mart{\'\i}nez}, {Masuda}, {Mazin},
  {Mi{\'c}anovi{\'c}}, {Miceli}, {Minev}, {Miranda}, {Mirzoyan}, {Molina},
  {Moralejo}, {Morcuende}, {Moreno}, {Moretti}, {Munar-Adrover}, {Neustroev},
  {Nigro}, {Nilsson}, {Ninci}, {Nishijima}, {Noda}, {Nogu{\'e}s}, {N{\"o}the},
  {Nozaki}, {Paiano}, {Palacio}, {Palatiello}, {Paneque}, {Paoletti},
  {Paredes}, {Pe{\~n}il}, {Peresano}, {Persic}, {Prada Moroni}, {Prand ini},
  {Puljak}, {Rhode}, {Rib{\'o}}, {Rico}, {Righi}, {Rugliancich}, {Saha},
  {Sahakyan}, {Saito}, {Sakurai}, {Satalecka}, {Schmidt}, {Schweizer},
  {Sitarek}, {{\v{S}}nidari{\'c}}, {Sobczynska}, {Somero}, {Stamerra}, {Strom},
  {Strzys}, {Suda}, {Suri{\'c}}, {Takahashi}, {Tavecchio}, {Temnikov},
  {Terzi{\'c}}, {Teshima}, {Torres-Alb{\`a}}, {Tosti}, {Tsujimoto}, {Vagelli},
  {van Scherpenberg}, {Vanzo}, {Vazquez Acosta}, {Vigorito}, {Vitale}, {Vovk},
  {Will}, {Zari{\'c}}, \& {Nava}}]{MAGIC-190114C-a}
{MAGIC Collaboration}, {Acciari}, V.~A., {Ansoldi}, S., {et~al.}
  2019{\natexlab{a}}, \nat, 575, 455, \dodoi{10.1038/s41586-019-1750-x}

\bibitem[{{MAGIC Collaboration} {et~al.}(2019{\natexlab{b}}){MAGIC
  Collaboration}, {Acciari}, {Ansoldi}, {Antonelli}, {Engels}, {Baack},
  {Babi{\'c}}, {Banerjee}, {Barres de Almeida}, {Barrio}, {Becerra
  Gonz{\'a}lez}, {Bednarek}, {Bellizzi}, {Bernardini}, {Berti}, {Besenrieder},
  {Bhattacharyya}, {Bigongiari}, {Biland }, {Blanch}, {Bonnoli},
  {Bo{\v{s}}njak}, {Busetto}, {Carosi}, {Ceribella}, {Chai}, {Chilingaryan},
  {Cikota}, {Colak}, {Colin}, {Colombo}, {Contreras}, {Cortina}, {Covino},
  {D'Elia}, {da Vela}, {Dazzi}, {de Angelis}, {de Lotto}, {Delfino}, {Delgado},
  {Depaoli}, {di Pierro}, {di Venere}, {Do Souto Espi{\~n}eira}, {Dominis
  Prester}, {Donini}, {Dorner}, {Doro}, {Elsaesser}, {Fallah Ramazani},
  {Fattorini}, {Ferrara}, {Fidalgo}, {Foffano}, {Fonseca}, {Font}, {Fruck},
  {Fukami}, {Garc{\'\i}a L{\'o}pez}, {Garczarczyk}, {Gasparyan}, {Gaug},
  {Giglietto}, {Giordano}, {Godinovi{\'c}}, {Green}, {Guberman}, {Hadasch},
  {Hahn}, {Herrera}, {Hoang}, {Hrupec}, {H{\"u}tten}, {Inada}, {Inoue},
  {Ishio}, {Iwamura}, {Jouvin}, {Kerszberg}, {Kubo}, {Kushida}, {Lamastra},
  {Lelas}, {Leone}, {Lindfors}, {Lombardi}, {Longo}, {L{\'o}pez},
  {L{\'o}pez-Coto}, {L{\'o}pez-Oramas}, {Loporchio}, {Machado de Oliveira
  Fraga}, {Maggio}, {Majumdar}, {Makariev}, {Mallamaci}, {Maneva}, {Manganaro},
  {Mannheim}, {Maraschi}, {Mariotti}, {Mart{\'\i}nez}, {Mazin},
  {Mi{\'c}anovi{\'c}}, {Miceli}, {Minev}, {Mirand a}, {Mirzoyan}, {Molina},
  {Moralejo}, {Morcuende}, {Moreno}, {Moretti}, {Munar-Adrover}, {Neustroev},
  {Nigro}, {Nilsson}, {Ninci}, {Nishijima}, {Noda}, {Nogu{\'e}s}, {Nozaki},
  {Paiano}, {Palatiello}, {Paneque}, {Paoletti}, {Paredes}, {Pe{\~n}il},
  {Peresano}, {Persic}, {Moroni}, {Prandini}, {Puljak}, {Rhode}, {Rib{\'o}},
  {Rico}, {Righi}, {Rugliancich}, {Saha}, {Sahakyan}, {Saito}, {Sakurai},
  {Satalecka}, {Schmidt}, {Schweizer}, {Sitarek}, {{\v{S}}nidari{\'c}},
  {Sobczynska}, {Somero}, {Stamerra}, {Strom}, {Strzys}, {Suda}, {Suri{\'c}},
  {Takahashi}, {Tavecchio}, {Temnikov}, {Terzi{\'c}}, {Teshima},
  {Torres-Alb{\`a}}, {Tosti}, {Vagelli}, {van Scherpenberg}, {Vanzo}, {Vazquez
  Acosta}, {Vigorito}, {Vitale}, {Vovk}, {Will}, {Zari{\'c}}, {Nava}, {Veres},
  {Bhat}, {Briggs}, {Cleveland }, {Hamburg}, {Hui}, {Mailyan}, {Preece},
  {Roberts}, {von Kienlin}, {Wilson-Hodge}, {Kocevski}, {Arimoto}, {Tak},
  {Asano}, {Axelsson}, {Barbiellini}, {Bissaldi}, {Dirirsa}, {Gill}, {Granot},
  {McEnery}, {Omodei}, {Razzaque}, {Piron}, {Racusin}, {Thompson}, {Campana},
  {Bernardini}, {Kuin}, {Siegel}, {Cenko}, {O'Brien}, {Capalbi}, {Da{\i}}, {de
  Pasquale}, {Gropp}, {Klingler}, {Osborne}, {Perri}, {Starling},
  {Tagliaferri}, {Tohuvavohu}, {Ursi}, {Tavani}, {Cardillo}, {Casentini},
  {Piano}, {Evangelista}, {Verrecchia}, {Pittori}, {Lucarelli}, {Bulgarelli},
  {Parmiggiani}, {Anderson}, {Anderson}, {Bernardi}, {Bolmer},
  {Caballero-Garc{\'\i}a}, {Carrasco}, {Castell{\'o}n}, {Castro Segura},
  {Castro-Tirado}, {Cherukuri}, {Cockeram}, {D'Avanzo}, {di Dato}, {Diretse},
  {Fender}, {Fern{\'a}ndez-Garc{\'\i}a}, {Fynbo}, {Fruchter}, {Greiner},
  {Gromadzki}, {Heintz}, {Heywood}, {van der Horst}, {Hu}, {Inserra}, {Izzo},
  {Jaiswal}, {Jakobsson}, {Japelj}, {Kankare}, {Kann}, {Kouveliotou}, {Klose},
  {Levan}, {Li}, {Lotti}, {Maguire}, {Malesani}, {Manulis}, {Marongiu},
  {Martin}, {Melandri}, {Micha{\l}owski}, {Miller-Jones}, {Misra}, {Moin},
  {Mooley}, {Nasri}, {Nicholl}, {Noschese}, {Novara}, {Pandey}, {Peretti},
  {P{\'e}rez Del Pulgar}, {P{\'e}rez-Torres}, {Perley}, {Piro}, {Ragosta},
  {Resmi}, {Ricci}, {Rossi}, {S{\'a}nchez-Ram{\'\i}rez}, {Selsing}, {Schulze},
  {Smartt}, {Smith}, {Sokolov}, {Stevens}, {Tanvir}, {Th{\"o}ne}, {Tiengo},
  {Tremou}, {Troja}, {de Ugarte Postigo}, {Valeev}, {Vergani}, {Wieringa},
  {Woudt}, {Xu}, {Yaron}, \& {Young}}]{MAGIC-190114C-b}
---. 2019{\natexlab{b}}, \nat, 575, 459, \dodoi{10.1038/s41586-019-1754-6}

\bibitem[{{MAGIC~Collaboration} {et~al.}(2020)}]{MAGIC-GRB13_15}
{MAGIC~Collaboration}, {et~al.} 2020, in prep.

\bibitem[{{Meegan} {et~al.}(2009){Meegan}, {Lichti}, {Bhat}, {Bissaldi},
  {Briggs}, {Connaughton}, {Diehl}, {Fishman}, {Greiner}, {Hoover}, {van der
  Horst}, {von Kienlin}, {Kippen}, {Kouveliotou}, {McBreen}, {Paciesas},
  {Preece}, {Steinle}, {Wallace}, {Wilson}, \& {Wilson-Hodge}}]{GBM-instrument}
{Meegan}, C., {Lichti}, G., {Bhat}, P.~N., {et~al.} 2009, \apj, 702, 791,
  \dodoi{10.1088/0004-637X/702/1/791}

\bibitem[{{Metzger}(2019)}]{Metzger19}
{Metzger}, B.~D. 2019, Living Reviews in Relativity, 23, 1,
  \dodoi{10.1007/s41114-019-0024-0}

\bibitem[{{Mirzoyan} {et~al.}(2019){Mirzoyan}, {Noda}, {Moretti}, {Berti},
  {Nigro}, {Hoang}, {Micanovic}, {Takahashi}, {Chai}, {Moralejo}, \& {MAGIC
  Collaboration}}]{MAGIC-190114C-GCN}
{Mirzoyan}, R., {Noda}, K., {Moretti}, E., {et~al.} 2019, GRB Coordinates
  Network, 23701, 1

\bibitem[{{Mooley} {et~al.}(2018){Mooley}, {Deller}, {Gottlieb}, {Nakar},
  {Hallinan}, {Bourke}, {Frail}, {Horesh}, {Corsi}, \& {Hotokezaka}}]{Mooley18}
{Mooley}, K.~P., {Deller}, A.~T., {Gottlieb}, O., {et~al.} 2018, \nat, 561,
  355, \dodoi{10.1038/s41586-018-0486-3}

\bibitem[{{Murase} {et~al.}(2011){Murase}, {Toma}, {Yamazaki}, \&
  {M{\'e}sz{\'a}ros}}]{Murase11}
{Murase}, K., {Toma}, K., {Yamazaki}, R., \& {M{\'e}sz{\'a}ros}, P. 2011, \apj,
  732, 77, \dodoi{10.1088/0004-637X/732/2/77}

\bibitem[{{Murase} {et~al.}(2018){Murase}, {Toomey}, {Fang}, {Oikonomou},
  {Kimura}, {Hotokezaka}, {Kashiyama}, {Ioka}, \&
  {M{\'e}sz{\'a}ros}}]{Murase18}
{Murase}, K., {Toomey}, M.~W., {Fang}, K., {et~al.} 2018, \apj, 854, 60,
  \dodoi{10.3847/1538-4357/aaa48a}

\bibitem[{{Nakar}(2019)}]{Nakar19}
{Nakar}, E. 2019, arXiv e-prints, arXiv:1912.05659.
\newblock \doarXiv{1912.05659}

\bibitem[{{Norris} \& {Bonnell}(2006)}]{Norris06}
{Norris}, J.~P., \& {Bonnell}, J.~T. 2006, \apj, 643, 266,
  \dodoi{10.1086/502796}

\bibitem[{Palmer {et~al.}(2016)}]{BATref-160821B}
Palmer, D.~M., {et~al.} 2016, GCN Circ., 19844

\bibitem[{{Panaitescu} \& {Kumar}(2000)}]{pk}
{Panaitescu}, A., \& {Kumar}, P. 2000, \apj, 543, 66, \dodoi{10.1086/317090}

\bibitem[{{Razzaque}(2010)}]{Razzaque10}
{Razzaque}, S. 2010, \apjl, 724, L109, \dodoi{10.1088/2041-8205/724/1/L109}

\bibitem[{{Rolke} {et~al.}(2005){Rolke}, {L{\'o}pez}, \& {Conrad}}]{Rolke}
{Rolke}, W.~A., {L{\'o}pez}, A.~M., \& {Conrad}, J. 2005, Nuclear Instruments
  and Methods in Physics Research A, 551, 493,
  \dodoi{10.1016/j.nima.2005.05.068}

\bibitem[{{Rossi} {et~al.}(2020){Rossi}, {Stratta}, {Maiorano}, {Spighi},
  {Masetti}, {Palazzi}, {Gardini}, {Melandri}, {Nicastro}, {Pian}, {Branchesi},
  {Dadina}, {Testa}, {Brocato}, {Benetti}, {Ciolfi}, {Covino}, {D'Elia},
  {Grado}, {Izzo}, {Perego}, {Piranomonte}, {Salvaterra}, {Selsing},
  {Tomasella}, {Yang}, {Vergani}, {Amati}, \& {Stephen}}]{Rossi20}
{Rossi}, A., {Stratta}, G., {Maiorano}, E., {et~al.} 2020, \mnras, 493, 3379,
  \dodoi{10.1093/mnras/staa479}

\bibitem[{{Sari} \& {Esin}(2001)}]{SariEsin}
{Sari}, R., \& {Esin}, A.~A. 2001, \apj, 548, 787, \dodoi{10.1086/319003}

\bibitem[{{Sari} {et~al.}(1998){Sari}, {Piran}, \& {Narayan}}]{sari98}
{Sari}, R., {Piran}, T., \& {Narayan}, R. 1998, \apjl, 497, L17,
  \dodoi{10.1086/311269}

\bibitem[{Siegel {et~al.}(2016)}]{Swift-160821B}
Siegel, M.~H., {et~al.} 2016, GCN Circ., 19833

\bibitem[{Stanbro {et~al.}(2016)Stanbro, Meegan, {et~al.}}]{GBM-160821B}
Stanbro, M., Meegan, C., {et~al.} 2016, GCN Circ., 19843

\bibitem[{{Tanvir} {et~al.}(2013){Tanvir}, {Levan}, {Fruchter}, {Hjorth},
  {Hounsell}, {Wiersema}, \& {Tunnicliffe}}]{Tanvir13}
{Tanvir}, N.~R., {Levan}, A.~J., {Fruchter}, A.~S., {et~al.} 2013, \nat, 500,
  547, \dodoi{10.1038/nature12505}

\bibitem[{Troja {et~al.}(2016)}]{HST-160821B}
Troja, E., {et~al.} 2016, GCN Circ., 20222

\bibitem[{{Troja} {et~al.}(2019){Troja}, {Castro-Tirado}, {Becerra
  Gonz{\'a}lez}, {Hu}, {Ryan}, {Cenko}, {Ricci}, {Novara},
  {S{\'a}nchez-R{\'a}mirez}, {Acosta-Pulido}, {Ackley}, {Caballero
  Garc{\'\i}a}, {Eikenberry}, {Guziy}, {Jeong}, {Lien}, {M{\'a}rquez}, {Pand
  ey}, {Park}, {Sakamoto}, {Tello}, {Sokolov}, {Sokolov}, {Tiengo}, {Valeev},
  {Zhang}, \& {Veilleux}}]{Troja19}
{Troja}, E., {Castro-Tirado}, A.~J., {Becerra Gonz{\'a}lez}, J., {et~al.} 2019,
  \mnras, 489, 2104, \dodoi{10.1093/mnras/stz2255}

\bibitem[{{Vietri}(1997)}]{Vietri97}
{Vietri}, M. 1997, \prl, 78, 4328, \dodoi{10.1103/PhysRevLett.78.4328}

\bibitem[{von Kienlin {et~al.}(2020)von Kienlin, Meegan, Paciesas, Bhat,
  Bissaldi, Briggs, Burns, Cleveland, Gibby, Giles, Goldstein, Hamburg, Hui,
  Kocevski, Mailyan, Malacaria, Poolakkil, Preece, Roberts, Veres, \&
  Wilson-Hodge}]{GBM-catalog}
von Kienlin, A., Meegan, C.~A., Paciesas, W.~S., {et~al.} 2020, The
  Astrophysical Journal, 893, 46, \dodoi{10.3847/1538-4357/ab7a18}

\bibitem[{{Vurm} \& {Beloborodov}(2017)}]{Vurm17}
{Vurm}, I., \& {Beloborodov}, A.~M. 2017, \apj, 846, 152,
  \dodoi{10.3847/1538-4357/aa7ddb}

\bibitem[{{Wang} {et~al.}(2001){Wang}, {Dai}, \& {Lu}}]{Wang01}
{Wang}, X.~Y., {Dai}, Z.~G., \& {Lu}, T. 2001, \apj, 556, 1010,
  \dodoi{10.1086/321608}

\bibitem[{{Woosley} \& {Bloom}(2006)}]{Woosley06}
{Woosley}, S.~E., \& {Bloom}, J.~S. 2006, \araa, 44, 507,
  \dodoi{10.1146/annurev.astro.43.072103.150558}

\bibitem[{Xu {et~al.}(2016)}]{NOT-160821B}
Xu, D., {et~al.} 2016, GCN Circ., 19834

\bibitem[{Zanin {et~al.}(2013)}]{MARS-2013}
Zanin, R., {et~al.} 2013, Proc. of 33rd ICRC

\bibitem[{{Zhang} \& {M{\'e}sz{\'a}ros}(2001)}]{Zhang01}
{Zhang}, B., \& {M{\'e}sz{\'a}ros}, P. 2001, \apj, 559, 110,
  \dodoi{10.1086/322400}

\bibitem[{{Zhang} {et~al.}(2018){Zhang}, {Lei}, {Zhang}, {Chen}, {Xiong}, \&
  {Song}}]{Zhang18}
{Zhang}, Q., {Lei}, W.~H., {Zhang}, B.~B., {et~al.} 2018, \mnras, 475, 266,
  \dodoi{10.1093/mnras/stx3229}

\end{thebibliography}





\end{document}